\documentclass[preprint,showpacs,preprintnumbers,amsmath,amssymb,nofootinbib]{revtex4}
\usepackage{graphicx,subfigure,textcomp}
\usepackage{dcolumn}
\usepackage{bm}
\usepackage{mathrsfs,slashed,amsmath}
\usepackage{natbib}
\usepackage{dsfont}
\allowdisplaybreaks
\begin{document}
\title{Chiral behavior of vector meson self energies}
\author{Peter C. Bruns}
\affiliation{Institut f\"ur Theoretische Physik, Universit\"at Regensburg, D-93040 Regensburg, Germany}
\author{Maximilian Emmerich}
\affiliation{Institut f\"ur Theoretische Physik, Universit\"at Regensburg, D-93040 Regensburg, Germany}
\author{Ludwig Greil}
\affiliation{Institut f\"ur Theoretische Physik, Universit\"at Regensburg, D-93040 Regensburg, Germany}
\author{Andreas Sch\"afer}
\affiliation{Institut f\"ur Theoretische Physik, Universit\"at Regensburg, D-93040 Regensburg, Germany}
\date{\today}
\begin{abstract}
We employ a chiral Lagrangian framework with three dynamical flavors to calculate the masses of the lowest-lying vector mesons to one loop accuracy, and use the resulting formulae to extrapolate recent QCDSF lattice data on vector meson masses and mass ratios to the physical point. Our representation for the vector meson self energies also enables us to discuss loop corrections to the $\omega\phi$ mixing amplitude.
\end{abstract}
\maketitle
\section{Introduction}
\label{sec:Intro}
Vector mesons have played a very important role in hadron physics from the early days on \cite{Frazer:1959gy,Sakurai:1960ju,Sakurai:1966zz,Weinberg:1967kj,Feynman:1973xc} and were theoretically studied using model Lagrangians for vector fields, or employing dispersion and/or current algebra techniques. They were sometimes also interpreted as gauge bosons of a hidden local symmetry \cite{Bando:1985rf,Bando:1984ej}. We refer to \cite{Meissner:1987ge} for a comprehensive review. Nowadays, a convenient tool to describe low-energy reactions and properties of hadrons is given by Chiral Perturbation Theory (ChPT) \cite{Weinberg:1978kz,Gasser:1983yg,Gasser:1984gg,Gasser:1987rb,Leutwyler:1993iq}, the low-energy effective field theory of Quantum Chromodynamics (QCD) (see e.~g. \cite{Bernard:2007zu} for a recent review). In this framework, the pions (and in the case of three dynamical quark flavors also the kaons and the eta meson) are considered as the Goldstone bosons of spontaneously broken chiral symmetry. The latter is an exact symmetry of the QCD Lagrangian when the light quark masses are set to zero and no electroweak interaction is present - a situation that was sometimes called a ``theoretical paradise'' \cite{Leutwyler:2012ax}. In the real world, the masses of the $u$, $d$ and the $s$ - quark are nonzero, but small compared to a typical hadronic scale of $\Lambda_{had}\sim 1$ GeV, while the ``heavy'' quarks ($c,b,t$) are not active as dynamical degrees of freedom (d.o.f.) and can be integrated out of the theory. Moreover, quarks and gluons are confined inside the hadrons, so that the long-range part of the strong interaction is dominated by the Goldstone boson dynamics. This situation allows an effective-field-theory treatment of the interactions among hadrons, where the light quark masses and Goldstone boson momenta are treated as small quantities compared to $\Lambda_{had}$. One has to write down the most general effective Lagrangian consistent with chiral symmetry and all other symmetries of the underlying field theory (QCD), and imposes a suitable power-counting scheme to order the perturbation series in a low-energy expansion in the small quantities (meson momenta, quark masses etc.). Vector mesons were included in ChPT at an early stage \cite{Ecker:1989yg,Ecker:1988te} as massive matter fields interacting with the light Goldstone bosons. However, when the massive particles appear in a loop graph, it becomes non-trivial to keep the low-energy power counting manifest due to the introduction of a new ``heavy'' mass scale (the vector meson mass in the chiral limit). This phenomenon was also observed when incorporating baryons in ChPT on the one-loop level \cite{Gasser:1987rb}. To solve this problem, and to preserve the usual low-energy power counting scheme, a ``heavy vector meson theory'' was designed \cite{Jenkins:1995vb,Bijnens:1997ni,Bijnens:1997rv}, while schemes preserving the power counting {\em and}\, manifest Lorentz covariance when including vector mesons were worked out some years later \cite{Fuchs:2003sh,Bruns:2004tj,Bruns:2008ub,Djukanovic:2009zn}. All these schemes face a problem in the resonance energy region, due to the fact that the $\rho$ vector meson is not a stable particle under the strong interaction and can decay into two light Goldstone bosons (pions) which, by energy-momentum conservation, cannot be both of ``soft'' momentum (this problem does not occur for baryons since a decay into Goldstone bosons is prohibited by baryon number conservation): the imaginary part of the loop diagram which generates the decay width of the vector meson does not scale as expected from the na\"{i}ve application of the low-energy counting rules to the diagram. In the language of the infrared regularization scheme \cite{Becher:1999he,Schindler:2003xv} this part should belong to the ``regular part'' of the loop integral, which is usually simply dropped in infrared regularization, with the argument that it only contains analytic terms which can be absorbed in the local operators of the effective Lagrangian. In the present case, however, it is in general complex, and contains relevant physics. This problem is discussed in \cite{Bruns:2004tj,Djukanovic:2009zn}. While it is argued in \cite{Djukanovic:2009zn,Bauer:2012gn} that the power-counting violating portion of the imaginary part can be absorbed in renormalized masses and couplings (which then become complex), without spoiling perturbative unitarity, this procedure is certainly only valid when the resonance mass is far above the decay threshold, e.g. $2M_{\pi}\ll M_{\rho}$. In the present contribution, we want to study the vector meson masses for three dynamical flavors, within a chiral Lagrangian framework, to one-loop accuracy, using lattice data from the QCDSF collaboration \cite{Bietenholz:2011qq}. We will see that for most of the data points, the above requirement given by the inequality is not met. Of course, one could object that in this case the ChPT treatment is not valid any more, and some model dependence is involved in the quark mass region where the vector mesons suddenly ``become stable''. We are aware of that matter and consider our study as an exploratory one, which, however, fully incorporates all the one-loop effects relevant for the vector meson masses, widths and mixing amplitudes. In conclusion, it seems that the usual low-energy power counting of meson ChPT is not adapted to the analysis of the physics we want to investigate here. We will explain our approach to this problem in Sect.~\ref{sec:ExtForm}. Note that the relevance of the non-analyticities due to resonance decay thresholds for chiral extrapolations was recently discussed in \cite{Guo:2011gc}.\\An alternative way of examining the properties of meson resonances, instead of explicitly including them as fields in some Lagrangian, is to study a scattering process (or form factor) where these resonances show up, using some model scattering amplitudes which obey two-particle unitarity. For example, one can use a convenient model amplitude for $\pi\pi$ scattering to examine the properties of the (modelled) $\rho$ resonance. Such ideas are more than fifty years old \cite{GellMann:1961tg,Brown:1968zza} and have been revived some time ago employing ``Unitarized ChPT'' in \cite{Dobado:1989qm,Oller:1998hw,Oller:1998zr,Nieves:1999bx,GomezNicola:2001as}. In \cite{Hanhart:2008mx,Nebreda:2010wv,Pelaez:2010fj,Nebreda:2011di}, the quark mass dependence of the $\rho$ and $\sigma$ resonance masses was studied within such an approach.  It would be very nice to see a consistent picture emerge when comparing the chiral Lagrangian framework to such non-perturbative approaches. However, it is not clear a priori that the subset of Feynman graphs that is effectively resummed in the unitarized scattering amplitudes is sufficient to generate the correct quark mass dependence of the resonance parameters. For example, when studying the quark mass dependence of the $\rho$ mass, one must take care that $M_{\pi}^{3}$ terms are included, which are nonanalytic in the quark masses and are generated by the $\omega\pi$ sunset graphs, but not by the purely pionic loops (see e.g. \cite{Bijnens:1997ni,Bruns:2004tj,Djukanovic:2009zn}). In this work, we will not make use of such non-perturbative methods, and restrict ourselves to the one-loop level of perturbation theory to study the quark mass dependence of the vector meson self-energies. For some earlier studies of vector meson self-energies on the one-loop level, outside the framework of ChPT, we refer to \cite{Komar:1960af,Rollnik:1962zz,Chiang:1973qh,Benayoun:2000ti}. \\This article is organized as follows: In Sect.~\ref{sec:Generalities}, we present and explain the general formalism needed to compute the one-particle propagators of the vector particles on the one-loop level. In Sect.~\ref{sec:ExtForm}, we calculate the relevant vertices and one-loop graphs, and in Sect.~\ref{sec:Results}, we present and discuss the results of our approach, and draw some conclusions regarding these results. Explicit expressions for the occuring loop integrals can be found in the appendices. 
\newpage
\section{General formalism}
\label{sec:Generalities}
The free Lagrangian for massive vector fields $V_{\mu}$, $S_{\mu}$  is given by
\begin{align}
\mathcal{L}^{V}_{\mathrm{free}} &= -\frac{1}{4}\langle V_{\mu\nu}V^{\mu\nu}\rangle + \frac{1}{2}M_{V,b}^{2}\langle V_{\mu}V^{\mu}\rangle,\label{eq:Lfree1}\\
\mathcal{L}^{S}_{\mathrm{free}} &= -\frac{1}{4} S_{\mu\nu}S^{\mu\nu} + \frac{1}{2}M_{S,b}^{2} S_{\mu}S^{\mu},\label{eq:Lfree2}
\end{align}
where $V_{\mu\nu}=\nabla_{\mu}V_{\nu}-\nabla_{\nu}V_{\mu}$ and $S_{\mu\nu}=\partial_{\mu}S_{\nu}-\partial_{\nu}S_{\mu}$ are the field strength tensors associated with the vector fields $V_{\mu}$, $S_{\mu}$. The brackets $\langle\ldots\rangle$ denote the trace in flavor space. The lowest-lying vector meson octet is contained in 
\begin{align}
V_{\mu} = V_{\mu}^{a}\lambda^{a} = \begin{pmatrix}\frac{\rho^{0}}{\sqrt{2}}+\frac{\phi^{(8)}}{\sqrt{6}} & \rho^{+}& K^{*+} \\ \rho^{-} & -\frac{\rho^{0}}{\sqrt{2}}+\frac{\phi^{(8)}}{\sqrt{6}} & K^{*0} \\
K^{*-} & \bar K^{*0}  & -\frac{2\phi^{(8)}}{\sqrt{6}}\end{pmatrix}_{\mu}.
\end{align}
We also introduce a singlet field $S_{\mu}=\phi_{\mu}^{(0)}$. The ``bare masses'' $M_{V/S,b}$ are interpreted as the masses of the vector fields when all interactions are turned off. We are only interested in the contributions to the self energy of the vector mesons due to the interaction with the lowest-lying octet of pseudoscalar mesons $\varphi$, which are interpreted as the pseudo-Goldstone-Bosons (PGBs) of spontaneously broken chiral $SU(3)_{L}\times SU(3)_{R}$ symmetry \cite{Gasser:1984gg}, collected in a matrix $U=u^{2}$,
\begin{align}
U=\exp\left(\frac{\sqrt{2}i\varphi}{F_{0}}\right),\quad \varphi=\varphi^{i}\lambda^{i}=\begin{pmatrix}\frac{\pi^{0}}{\sqrt{2}}+\frac{\eta}{\sqrt{6}}& \pi^{+} & K^{+} \\ \pi^{-}& -\frac{\pi^{0}}{\sqrt{2}}+\frac{\eta}{\sqrt{6}}& K^{0} \\K^{-}& \bar K^{0} & -\frac{2\eta}{\sqrt{6}} \end{pmatrix}.
\end{align}
The interaction Lagrangians needed for the calculation of the vector meson self-energies are fairly standard by now \cite{Ecker:1989yg,Ecker:1988te,Jenkins:1995vb,Urech:1995ry,Birse:1996hd,Klingl:1996by,Cirigliano:2003yq,Rosell:2004mn,Lutz:2008km,Leupold:2012qn,Terschlusen:2013iqa} (see also \cite{Armour:2005mk,Grigoryan:2005zj} for the 'partially quenched' case): There is a term linear in the vector fields, describing e.g. the decay vertex $\rho\rightarrow\pi\pi$,
\begin{align}\label{eq:Llin}
\mathcal{L}_{\mathrm{lin}}= -\frac{ig_{V}}{2\sqrt{2}}\langle\lbrack u_{\mu},u_{\nu}\rbrack V^{\mu\nu}\rangle,
\end{align}
and there are also bilinear terms,
\begin{align}\label{eq:Lnonlin}
\mathcal{L}_{VV\phi}= \frac{g_{A}^{V}}{2}\epsilon_{\mu\nu\rho\sigma}\langle\lbrace\nabla^{\mu}V^{\nu},V^{\rho}\rbrace u^{\sigma}\rangle + g_{A}^{VS}\epsilon_{\mu\nu\rho\sigma}\langle\nabla^{\mu}V^{\nu}S^{\rho}u^{\sigma}\rangle.
\end{align}
We note that the $V\rightarrow \varphi\varphi,\,V\rightarrow V\varphi$ and $S\rightarrow V\varphi$ vertizes derived from the above Lagrangians are transversal in the sense that the contraction of the vertex rules with the four-momentum $k^{\mu}$ of a vector field vanishes. Thus the scalar degrees of freedom of the four-vector fields decouple from the PGBs. For a review of the problems with additional degrees of freedom, and other general aspects in the description of spin-1 fields in an effective field theory framework, we refer to the recent study in \cite{Kampf:2009jh}. We also note that we set the external source fields $\hat{v}_{\mu},\hat{a}_{\mu},\hat{p}$, introduced in the general ChPT framework \cite{Gasser:1984gg,Ecker:1988te}, to zero. The correct explicit chiral symmetry breaking known from QCD is implemented by coupling the effective fields to an external matrix source field $\hat{s}(x)$, which is set equal to the quark mass matrix $\mathcal{M}=$\,$\mathrm{diag}(m_{\ell},m_{\ell},m_{s})$ in the end. We use the notation familiar from ChPT \cite{Gasser:1984gg,Ecker:1988te},
\begin{align}
\begin{split}
u_{\mu} &= iu^{\dagger}(\nabla_{\mu}U)u^{\dagger},\quad \chi_{+} = u^{\dagger}\chi u^{\dagger} + u\chi^{\dagger}u,\quad \chi = 2B_{0}\hat{s}\rightarrow 2B_{0}\mathcal{M},\\
\nabla_{\mu}V_{\nu} &= \partial_{\mu}V_{\nu} + \lbrack\Gamma_{\mu},V_{\nu}\rbrack,\quad \Gamma_{\mu} = \frac{1}{2}\left(u^{\dagger}\partial_{\mu}u + u\partial_{\mu}u^{\dagger}\right). 
\end{split}
\end{align}
We neglect isospin breaking effects and set $m_{u}=m_{d}=m_{\ell}$. $F_{0}$ is the PGB decay constant in the three-flavor chiral limit $m_{\ell,s}\rightarrow 0$, while the constant $B_{0}$ is proportional to the quark condensate in the same limit \cite{Gasser:1984gg}. The contact term Lagrangian including the source field $\hat{s}(x)$ will be given below.
\begin{figure}
\includegraphics[width=0.3\textwidth]{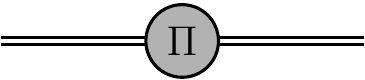}
\caption{The vector meson self energy $\Pi_{\mu\nu}(k)$. The double line stands for the incoming/outgoing vector meson.}
\label{fig:Pi}
\end{figure}
\subsection{One-particle propagator}
\label{subsec:Propagator}
The free propagator of the vector field, in momentum space, is derived from Eq.~(\ref{eq:Lfree1}) (setting $s=k^{2}$)
\begin{align}\label{eq:freeprop1}
(\mathbf{D}_{0}^{-1})_{\mu\nu}(k) = \frac{(-i)\left(g_{\mu\nu}-\frac{k_{\mu}k_{\nu}}{M_{V,b}^{2}}\right)}{s-M_{V,b}^{2}} = \frac{(-i)\left(g_{\mu\nu}-\frac{k_{\mu}k_{\nu}}{k^{2}}\right)}{s-M_{V,b}^{2}} + i\frac{k_{\mu}k_{\nu}}{k^{2}M_{V,b}^{2}}.
\end{align}
Note that we assume $M_{V,b}$ to be {\em real}\,: The width of the vector meson resonance is generated by the dressing due to meson loops. We split up the self energy (see Fig.~\ref{fig:Pi}) in a transversal and a longitudinal part,
\begin{align}
\Pi_{\mu\nu}(k) = \left(g_{\mu\nu}-\frac{k_{\mu}k_{\nu}}{k^{2}}\right)\Pi_{T}(s)+\frac{k_{\mu}k_{\nu}}{k^{2}}\Pi_{L}(s)
\end{align}
and resum the geometric series of two-point graphs
\begin{align}
\begin{split}
(\mathbf{D}^{-1})_{\mu\nu}(k) &= (\mathbf{D}_{0}^{-1})_{\mu\nu}+(\mathbf{D}_{0}^{-1})_{\mu\alpha}(i\Pi)^{\alpha\beta}(\mathbf{D}_{0}^{-1})_{\beta\nu}+\ldots = \left(\mathbf{D}_{0} - i\mathbf{\Pi}\right)^{-1}_{\mu\nu}. 
\end{split}
\end{align}
One easily finds
\begin{align}
(\mathbf{D}_{0})_{\mu\nu} = i((k^{2}-M_{V,b}^{2})g_{\mu\nu}-k_{\mu}k_{\nu}),
\end{align}
and by matrix inversion
\begin{align}\label{eq:fullprop1}
(\mathbf{D}^{-1})_{\mu\nu}(k) = \frac{(-i)\left(g_{\mu\nu}-\frac{k_{\mu}k_{\nu}}{k^{2}}\right)}{s-M_{V,b}^{2}-\Pi_{T}(s)} + \frac{ik_{\mu}k_{\nu}}{k^{2}(M_{V,b}^{2}+\Pi_{L}(s))}.
\end{align}
The first term is of a form similar to the transversal (i.e. spin 1) part of the free propagator, with a pole position shifted perturbatively by $\Pi_{T}$, while the second term does not contain a pole in the vicinity of $M_{V,b}^{2}$, given that perturbation theory is reliable here. Moreover, the second term drops out when it is dotted between the vertices from Eq.~(\ref{eq:Llin}), due to the transversality property mentioned above. We therefore concentrate on the calculation of $\Pi_{T}$, but note that one should have $\Pi_{T}(0) = \Pi_{L}(0)$ for general interactions, to assure that the self energy does not have a pole at $s\equiv k^{2}=0$. Moreover, we are only interested in the contribution to $\Pi_{T}$ which is due to the interaction with the PGBs. In the following, we will assume that all other hadronic contributions have already been absorbed in the parameters occuring there, and in $M_{V,b}^{2}$. This is permissible in an effective field theory treatment designed for the description of low energy interactions. Let us first treat the chiral limit case, where $\mathcal{M}\rightarrow 0$. Then the denominator of the transversal part of the full propagator reads
\begin{align}\label{eq:denomchlim}
s-M_{V,b}^{2}-\mathring{\Pi}_{T}^{PGB}(s) = s-M_{V,b}^{2}-\mathring{\Pi}_{T,loop}^{PGB}(s) - \sum_{n=0}^{N}\mathring{c}_{n}s^{n},
\end{align}
where the $\circ$ denotes the function in the chiral limit. The $c_{n}$-terms are counterterms needed to absorb the divergences in the ``$loop$'' part. $N$ depends on the degree of divergence of the loop graphs. It is straightforward to construct the corresponding counterterm Lagrangian for such energy-dependent terms, see e.g. \cite{Rosell:2004mn,Kampf:2009jh}. We will not need the explicit form of these terms here.\\Since the relevant interaction vertices of the vector mesons with the PGBs (from Eqs.~(\ref{eq:Llin},\ref{eq:Lnonlin})) share the transversality property, we can set the longitudinal part of the self energy to zero w.l.o.g., $\mathring{\Pi}_{L}^{PGB}(s)=0$, so that we must also have $\mathring{\Pi}_{T}^{PGB}(0)=0$. In general, local couplings contributing only to $\mathring{\Pi}_{L}$ can be transformed away by a field redefinition affecting only the scalar (longitudinal, spin-0) component of the four-vector field (compare also the remarks on Eq.~(3.9) in \cite{Ecker:1988te}). Therefore we should have
\begin{align}\label{eq:nolongPi}
\mathring{c}_{0}=0\qquad\mathrm{and}\qquad \mathring{\Pi}_{T,loop}^{PGB}(s)\overset{s\rightarrow 0}{\longrightarrow} 0.
\end{align}
The bare mass $M_{V,b}$ is thus not renormalized by the loop graphs calculated in this work.\\We define two real parameters $\mathring{M}_{V}$ and $\mathring{\Gamma}_{V}$ to denote the pole position of the propagator in the chiral limit, $\mathring{s}_{pole}=\mathring{M}_{V}^{2}-i\mathring{M}_{V}\mathring{\Gamma}_{V}$, so that
\begin{align}\label{eq:polchlim}
\mathring{s}_{pole}-M_{V,b}^{2}-\left(\mathring{\Pi}_{T,loop}^{PGB}(\mathring{s}_{pole}) + \sum_{n=1}^{N}\mathring{c}_{n}\mathring{s}_{pole}^{n}\right)=0.
\end{align}
In this work, the corrections in the round brackets will be treated only to one-loop accuracy. Moreover, we assume that the width $\mathring{\Gamma}_{V}$ is sufficiently small compared to $\mathring{M}_{V}$,
\begin{align}\label{eq:smallwidth}
\frac{\mathring{\Gamma}_{V}^{2}}{\mathring{M}_{V}^{2}} \ll 1,
\end{align}
so that we can neglect terms of quadratic order in the imaginary part of the pole position. The validity of this assumption will be discussed later. 
In addition, the vector field propagators occuring in some of the loop functions are taken as the free propagators (see Eq.~(\ref{eq:freeprop1})) with a pole position shifted to $\mathring{s}_{pole}$, so that the width can immediately be related to the imaginary part of the loop integrals occuring in $\mathring{\Pi}_{T,loop}^{PGB}(\mathring{M}_{V}^{2})$ within these approximations,
\begin{align}\label{eq:oneloopwidth}
\mathring{M}_{V}\mathring{\Gamma}_{V} = -\mathrm{Im}\,\mathring{\Pi}_{T,loop}^{PGB}(\mathring{s}_{pole}) -\mathrm{Im}\,\sum_{n=1}^{N}\mathring{c}_{n}\mathring{s}_{pole}^{n} \approx -\mathrm{Im}\,\mathring{\Pi}_{T,loop}^{PGB}(\mathring{M}_{V}^{2}).
\end{align}
Obviously, the difference between $\mathring{M}_{V}^{2}$ and $M_{V,b}^{2}$, and also the width $\mathring{\Gamma}_{V}$, amounts to a two-loop effect when inserted in the loop corrections. Then, we can eliminate the unobservable parameter $M_{V,b}^{2}$ to one-loop accuracy,
\begin{align}\label{eq:polchlimOneLoop}
\begin{split}
M_{V,b}^{2} &= \mathring{M}_{V}^{2}-\mathrm{Re}\,\mathring{\Pi}_{T,loop}^{PGB}(\mathring{s}_{pole}) - \mathrm{Re}\sum_{n=1}^{N}\mathring{c}_{n}\mathring{s}_{pole}^{n}\\ 
&\equiv \mathring{M}_{V}^{2}-\mathrm{Re}\,\mathring{\Pi}_{T,loop}^{PGB}(\mathring{s}_{pole}) - \mathrm{Re}\sum_{n=0}^{N}\mathring{d}_{n}(\mathring{s}_{pole}-\mathring{M}_{V}^{2})^{n}\\
&\approx \mathring{M}_{V}^{2}-\mathrm{Re}\,\mathring{\Pi}_{T,loop}^{PGB}(\mathring{s}_{pole}) - \mathring{d}_{0}, 
\end{split}
\end{align}
where we have used the approximation indicated in Eq.~(\ref{eq:smallwidth}). We stress that we rely here on the applicability of perturbation theory, but not on the convergence of the (low-energy) expansion in $s$. Indeed, being interested in the resonance region, it is appropriate to reorder the series of counterterms $c_{n}$ like 
\begin{align}\label{eq:reordering}
\sum_{n=0}^{N}c_{n}s^{n}=\sum_{n=0}^{N}d_{n}(s-\mathring{M}_{V}^{2})^{n},\qquad \sum_{n=0}^{N}(-1)^{n}\mathring{d}_{n}\mathring{M}_{V}^{2n}=0.
\end{align}
Expanding the denominator of Eq.~(\ref{eq:denomchlim}) around $\mathring{s}_{pole}$, using Eqs.~(\ref{eq:polchlimOneLoop},\ref{eq:reordering}), one finds, neglecting terms of $\mathcal{O}((s-\mathring{s}_{pole})^{2})$,
\begin{align}\label{eq:chlimpropnearpole}
\begin{split}
s-M_{V,b}^{2}-\mathring{\Pi}_{T}^{PGB}(s) &\approx  \mathring{s}_{pole} + (s-\mathring{s}_{pole})\left(1-\frac{d}{ds}\mathring{\Pi}_{T,loop}^{PGB}|_{s=\mathring{s}_{pole}} - \sum_{n=1}^{N}n\mathring{d}_{n}(\mathring{s}_{pole}-\mathring{M}_{V}^{2})^{n-1}\right)\\
&\quad- \left(\mathring{M}_{V}^{2}+\mathring{\Pi}_{T,loop}^{PGB}(\mathring{s}_{pole})-\mathrm{Re}\,\mathring{\Pi}_{T,loop}^{PGB}(\mathring{s}_{pole}) + i\mathrm{Im}\sum_{n=1}^{N}\mathring{d}_{n}(\mathring{s}_{pole}-\mathring{M}_{V}^{2})^{n}\right)\\ 
&\approx  (s-\mathring{s}_{pole})\left(1-\frac{d}{ds}\mathring{\Pi}_{T,loop}^{PGB}|_{s=\mathring{s}_{pole}}-\mathring{d}_{1} + 2i\mathring{d}_{2}\mathring{M}_{V}\mathring{\Gamma}_{V}\right)\\ 
&\quad+ \mathring{s}_{pole}-\left(\mathring{M}_{V}^{2}-i\mathring{M}_{V}\mathring{\Gamma}_{V}\right). 
\end{split}
\end{align}
In the vicinity of the pole, the transversal part of the propagator is therefore of the form
\begin{align}\label{eq:transfullprop}
(\mathring{\mathbf{D}}^{-1})_{\mu\nu}^{T}(k) = \frac{(-i)\mathring{R}\left(g_{\mu\nu}-\frac{k_{\mu}k_{\nu}}{k^{2}}\right)}{s-\mathring{s}_{pole}} ,
\end{align}
where the residue $\mathring{R}$ can be read from Eq.~(\ref{eq:chlimpropnearpole}).
We shall require the renormalization condition $\mathrm{Re}\,R=1$ (see also Sect.~5 of \cite{Klingl:1996by}), so that to one-loop order we must have 
\begin{align}\label{eq:rescond}
 \mathring{d}_{1}\overset{!}{=}-\mathrm{Re}\,\frac{d}{ds}\mathring{\Pi}_{T,loop}^{PGB}|_{s=\mathring{s}_{pole}}.
\end{align}
On the basis of this treatment of the chiral limit case, we find in the general case ($\mathcal{M}\not= 0$)
\begin{align}
s-M_{V,b}^{2}-\Pi_{T}^{PGB}(s) &= s - \biggl(\mathring{M}_{V}^{2} + \bar{\Pi}_{T,loop}^{PGB}(s) + e_{0} +\sum_{n=1}^{N}d_{n}(s-\mathring{M}_{V}^{2})^{n} - D_{N}\biggr), \label{eq:selfEfull}\\
\bar{\Pi}_{T,loop}^{PGB}(s) &= \Pi_{T,loop}^{PGB}(s) - \mathrm{Re}\,\mathring{\Pi}_{T,loop}^{PGB}(\mathring{s}_{pole}), \label{def:PiBar}\\
D_{N} &= \mathrm{Re}\sum_{n=1}^{N}\mathring{d}_{n}(\mathring{s}_{pole}-\mathring{M}_{V}^{2})^{n} = \mathrm{Re}\sum_{n=2}^{N}\mathring{d}_{n}(-i\mathring{M}_{V}\mathring{\Gamma}_{V})^{n},\label{def:DN}\\
e_{n} &= d_{n}-\mathring{d}_{n}.\label{def:en}
\end{align}
The $e_{n}$ terms contain contributions from quark mass dependent counter\-terms. Note that $M_{V,b}$ in Eq.~(\ref{eq:selfEfull}) is the same parameter as in Eq.~(\ref{eq:polchlim}), because the quark mass corrections to the bare mass are treated as a further perturbation (in addition to the PGB loops), and are given at leading order by $e_{0}$. Also note that the constant $D_{N}$ is of two-loop order (of order $\mathring{\Gamma}_{V}^{2}$); it is neglected in our application of the above formulae.\\Consider the pole position $s_{pole}$ of the propagator in the case of non-vanishing quark masses. Examining the relevant loop graphs it turns out that the first corrections non-analytic in the quark masses are of $\mathcal{O}(m_{q}^{3/2})\sim\mathcal{O}(M_{PGB}^{3})$, while the quark mass dependent counterterms yield only even powers of $M_{PGB}$. Schematically,
\begin{align}
s_{pole}&=\mathring{s}_{pole}+x_{2}M_{PGB}^{2}+x_{3}M_{PGB}^{3}+\mathcal{O}(M_{PGB}^{4}\log M_{PGB},M_{PGB}^{4}),\label{eq:spoleexp1}\\
e_{n} &= e_{n}^{(2)}M_{PGB}^{2}+\mathcal{O}(M_{PGB}^{4}).\label{enexp}
\end{align}
On expansion in $s_{pole}-\mathring{s}_{pole}$, one finds
\begin{align}
\begin{split}
0 &= s_{pole} - \biggl(\mathring{M}_{V}^{2} + \bar{\Pi}_{T,loop}^{PGB}(s_{pole}) + e_{0} +\sum_{n=1}^{N}d_{n}(s_{pole}-\mathring{M}_{V}^{2})^{n} - D_{N} \biggr)\\
  &= (s_{pole}-\mathring{s}_{pole}) -i\mathring{M}_{V}\mathring{\Gamma}_{V} - \mathrm{Re}\,\bar{\Pi}_{T,loop}^{PGB}(\mathring{s}_{pole}) -  \mathrm{Re}\sum_{n=0}^{N}e_{n}(\mathring{s}_{pole}-\mathring{M}_{V}^{2})^{n}\\ 
  &\quad- i\left(\mathrm{Im}\,\Pi_{T,loop}^{PGB}(\mathring{s}_{pole}) + \mathrm{Im}\sum_{n=1}^{N}d_{n}(\mathring{s}_{pole}-\mathring{M}_{V}^{2})^{n}\right)\\ 
  &\quad- (s_{pole}-\mathring{s}_{pole})\left(\frac{d}{ds}\Pi_{T,loop}^{PGB}|_{s=\mathring{s}_{pole}} + \sum_{n=1}^{N}nd_{n}(\mathring{s}_{pole}-\mathring{M}_{V}^{2})^{n-1}\right) + \mathcal{O}((s_{pole}-\mathring{s}_{pole})^{2})\\
  &\approx  (s_{pole}-\mathring{s}_{pole})\left(1-\frac{d}{ds}\Pi_{T,loop}^{PGB}|_{s=\mathring{s}_{pole}} - d_{1} + 2id_{2}\mathring{M}_{V}\mathring{\Gamma}_{V}\right)\\ 
  &\quad- \mathrm{Re}\,\bar{\Pi}_{T,loop}^{PGB}(\mathring{s}_{pole}) - e_{0} - i\mathrm{Im}\left(\Pi_{T,loop}^{PGB}(\mathring{M}_{V}^{2}) - \mathring{\Pi}_{T,loop}^{PGB}(\mathring{M}_{V}^{2})\right),\label{eq:spoleexp2} 
\end{split}
\end{align}
or, due to the condition of Eq.~(\ref{eq:rescond}), and Eqs.~(\ref{eq:spoleexp1},\ref{enexp}),
\begin{align}\label{eq:polposfull}
\begin{split}
s_{pole}-\mathring{s}_{pole} &= \mathrm{Re}\,\bar{\Pi}_{T,loop}^{PGB}(\mathring{s}_{pole}) + e_{0}^{(2)}M_{PGB}^{2} + i\mathrm{Im}\left(\Pi_{T,loop}^{PGB}(\mathring{M}_{V}^{2}) - \mathring{\Pi}_{T,loop}^{PGB}(\mathring{M}_{V}^{2})\right)\\ 
&\quad+ \mathcal{O}\left(M_{PGB}^{4}\log M_{PGB},M_{PGB}^{4}\right) + \mathcal{O}(\hbar^{2}),
\end{split}
\end{align}
where the last symbol stands for the two-loop terms neglected in the approximations indicated above. The coefficient $e_{1}^{(2)}$ can be fixed by the condition that $\mathrm{Re}\,R=1+\mathcal {O}\left(M_{PGB}^{4}\right)$. From the quark mass dependence of the vector meson masses and the above renormalization condition, we can fix $e_{0}$ and $d_{1}$ up to and including $\mathcal{O}\left(M_{PGB}^{2}\right)$. 
 As far as we are aware, there is no model-independent or natural way to determine the coefficients $\mathring{d}_{n\geq 2}$. Moreover, one has to be aware of the fact that the off-shell behavior of an amplitude like $\Pi_{T}(s)$ will in general depend on the chosen parameterization of the fields, see e.~g. \cite{Fearing:1999fw}.\\The expansion around the chiral limit $m_{\ell},m_{s}\sim M_{PGB}^{2}\rightarrow 0$ has some shortcomings: First, in the real world, the PGBs are not all light degrees of freedom compared to the vector meson masses, e.g. $\frac{M_{K}}{M_{K^{\ast}}}\sim 0.6$, $\frac{M_{\eta}}{M_{\omega}}\sim 0.7$, due to the large strange quark mass. Therefore the extrapolation from the chiral limit to the physical point is probably not under sufficient theoretical control (for a discussion of this point, for the case of baryon masses, see e.g. \cite{Bruns:2012eh} and references therein). Second, it is not a priori clear that a one-loop calculation will be sufficient for a faithful representation of the self energy close to the chiral limit, where the vector mesons can decay into states with three, four\ldots nearly massless PGBs. And third, the effects due to terms of $\mathcal{O}(\mathring{\Gamma}_{V}^{2}/\mathring{M}_{V}^{2})$ neglected in some intermediate approximations need not be tiny (as e.g. $\Gamma_{\rho}/M_{\rho}\sim 0.2$).\\In \cite{Bietenholz:2011qq,Bruns:2012eh} a different extrapolation to the physical point was explored, where the average quark mass $\bar{m}=\frac{1}{3}(2m_{\ell}+m_{s})$ was kept fixed at its physical value, while the flavor-SU(3) symmetry breaking combination $\delta m_{\ell}=m_{\ell}-\bar{m}=\frac{1}{3}(m_{\ell}-m_{s})$ was varied from zero to the physical value. It was argued in \cite{Bietenholz:2011qq} that this extrapolation method was of some advantage because the terms linear in $\delta m_{\ell}$ dominate the quark-mass dependence of the hadron masses for fixed $\bar{m}$ in a sufficiently broad region of the $(m_{\ell},m_{s})$ parameter space around the ``symmetric point'' where $\delta m_{\ell}=0$ and $\bar{m}=\bar{m}^{\mathrm{phys}}$. Let all quantities evaluated at this symmetric point be indexed with a star $\star$ (instead of the $\circ$ denoting the evaluation at the chiral limit where $\delta m_{\ell}=0$ {\em and}\, $\bar{m}=0$). E.g. the eight PGBs (PGB$=\lbrace\pi,K,\eta\rbrace$) are all of the same mass at the symmetric point,
\begin{align}
M_{PGB}^{2}(\delta m_{\ell}=0) = M_{\star}^{2} = 2B_{0}\bar{m}+\mathcal{O}(\bar{m}^{2}\log\bar{m}) \approx (412\,\mathrm{MeV})^{2}\quad\mathrm{for}\quad\bar{m}=\bar{m}^{\mathrm{phys}}.
\end{align}
According to the evaluation on the lattice presented in \cite{Bietenholz:2011qq}, the octet vector meson mass at the symmetric point is $M_{V}^{\star}(\bar{m}^{\mathrm{phys}})\approx 855\,\mathrm{MeV}$. Consequently, the octet vector mesons are almost stable particles there, which is certainly not a disadvantage when taking the symmetric point as a reference point instead of the chiral limit. The above equations (\ref{eq:selfEfull},\ref{eq:polposfull}) will {\em only} be used to analyze the running of $(s_{pole}^{\star}-\mathring{s}_{pole})$ when $\bar{m}$ is varied from $0$ to $\bar{m}^{\mathrm{phys}}$. The analysis of the symmetry breaking effects, including singlet-octet mixing, will make use of the reference point $\star$.\\To see how this works, reconsider Eq.~(\ref{eq:selfEfull}) and use (the first line of) Eq.~(\ref{eq:spoleexp2}) to write
\begin{align}
s_{pole}^{\star} = \mathring{M}_{V}^{2} + \bar{\Pi}_{T,loop}^{\star PGB}(s_{pole}^{\star}) + e_{0}^{\star} +\sum_{n=1}^{N}d_{n}^{\star}(s_{pole}^{\star}-\mathring{M}_{V}^{2})^{n} - D_{N}
\end{align}
and hence we find
\begin{align}
\begin{split}
s-M_{V,b}^{2}-\Pi_{T}^{PGB}(s) &= s - s_{pole}^{\star} - \left(\Pi_{T,loop}^{PGB}(s)-\Pi_{T,loop}^{\star PGB}(s_{pole}^{\star}) + e_{0} - e_{0}^{\star}\right)\\
                               &\quad- \left(\sum_{n=1}^{N}\delta d_{n}(s_{pole}^{\star}-\mathring{M}_{V}^{2})^{n} + \sum_{n=1}^{N}f_{n}(s-s_{pole}^{\star})^{n}\right),
\end{split}\label{eq:reorderingstep}\\
\delta d_{n} &= d_{n}-d_{n}^{\star},\\
f_{n} &= \sum_{m=n}^{N}d_{m}\binom{m}{n}(s_{pole}^{\star}-\mathring{M}_{V}^{2})^{m-n}.
\end{align}
In the following, we will neglect the tiny width at the symmetric point and set $\mathrm{Im}\,s_{pole}^{\star}=0$, $\mathrm{Re}\,s_{pole}^{\star} = (M_{V}^{\star})^{2}$. Since $\delta d_{n}=\mathcal{O}(\delta m_{\ell})$, $s_{pole}^{\star}-\mathring{M}_{V}^{2}=\mathcal{O}(\bar{m})$, the $\delta d_{n}$ terms give $\bar{m}$ corrections to the symmetry breaking terms in $e_{0} - e_{0}^{\star}$ and can be absorbed in the latter combination,
\begin{align}
\delta e = e_{0} - e_{0}^{\star} + \sum_{n=1}^{N}\delta d_{n}(s_{pole}^{\star}-\mathring{M}_{V}^{2})^{n} = \mathcal{O}(\delta m_{\ell}),
\end{align}
so that the final expression for the denominator of the vector meson propagator reads
\begin{align}\label{eq:newselfenergy}
s-M_{V,b}^{2}-\Pi_{T}^{PGB}(s) =  s - s_{pole}^{\star} - \left(\delta e + \Pi_{T,loop}^{PGB}(s)-\Pi_{T,loop}^{\star PGB}(s_{pole}^{\star}) + \sum_{n=1}^{N}f_{n}(s-s_{pole}^{\star})^{n}\right).
\end{align}
Since we fix $s_{pole}^{\star} = (M_{V}^{\star})^{2}$ from the lattice data, any reference to the chiral limit mass parameter $\mathring{M}_{V}$ has disappeared from Eq.~(\ref{eq:newselfenergy}). Also, for the fixed value of $\bar{m}$, the propagators in the loop functions are taken as free propagators with a pole position shifted to $M_{V}^{\star}$. The original series of counterterms has been reordered, trading the $d_{n}$ for the new coefficient functions $f_{n}(\bar{m},\delta m_{\ell})$, so the energy dependence is expanded around $M_{V}^{\star}$ instead of $\mathring{M}_{V}$. Of course, in the singlet case, one expands around $M_{S}^{\star}$ in complete analogy to the above. Moreover, we again require that the real part of the residue of the propagator at the pole is equal to one, in analogy to Eq.~(\ref{eq:rescond}), which determines the coefficient $f_{1}$ order by order in $\delta m_{\ell}$.
\subsection{Singlet-octet-mixing}
\label{subsec:Mixing}
In the general case $m_{\ell}\not=m_{s}$, the neutral octet isosinglet field not only mixes with $\varphi\varphi$ and $V\varphi,S\varphi$ states, but also with one-particle singlet states. This results in an additional complication: the one-particle propagator is non-diagonal in the $\phi^{(0)}-\phi^{(8)}$ sector \cite{Okubo:1963fa}. Hence, one has to invert the corresponding matrix
\begin{align}\label{eq:08matrix}
\mathbf{D}^{-1}_{\text{mix}}&=i\begin{pmatrix}s-M_{S}^{\star 2}-\delta\Pi_{00} & -\Pi_{08}\\-\Pi_{80} & s-M_V^{\star 2}-\delta\Pi_{88}\end{pmatrix}^{-1}\\
&=\frac{i}{\det_{08}}\begin{pmatrix}s-M_{V}^{\star 2}-\delta\Pi_{88} & \Pi_{08}\\\Pi_{80} & s-M_{S}^{\star 2}-\delta\Pi_{00}\end{pmatrix},
\end{align}
where again we only consider the transversal parts of the self-energies and propagators, and expand around the reference point $\star$. 
The expressions $\delta\Pi_{\ldots}$ thus stand for the differences $\Pi_{\ldots}(s)-\Pi_{\ldots}(s_{V,S,pole}^{\star})$ which also appear in Eq.~(\ref{eq:newselfenergy}), and where, respectively, $s_{V,S,pole}^{\star}=M_{V,S}^{\star 2}$ (here we neglect the tiny widths at the symmetric point as already noted above). In the case of the mixing amplitude $\Pi_{08}(s)$ the subtraction of course vanishes because there is no mixing at $\star$. Note that we use the notation $\Pi_{00}=\Pi_{\phi^{(0)}}$ and $\Pi_{88}=\Pi_{\phi^{(8)}}$ here for a better legibility, and to clarify the matrix notation.\\Since we are looking for the mass eigenvalues of the $\phi^{(0)}-\phi^{(8)}$ sector, we determine the (complex) zeros of the determinant $\det_{08}$ (see e.~g. Sect.~3 of \cite{Benayoun:2000ti}),
\begin{align}\label{eq:08det}
\text{det}_{08}=(s-M_V^{\star 2}-\delta\Pi_{88})(s-M_{S}^{\star 2}-\delta\Pi_{00})-\Pi_{08}\Pi_{80}.
\end{align}
In the simplified case of energy-independent self-energies $\Pi$, this results in a quadratic equation, the two roots of which are identified with the mass of the $\phi(1020)$ and the $\omega(782)$. This leads to the expressions for masses and the mixing angle $\Theta_{V}$ given e.~g. in \cite{Jenkins:1995vb}.
\section{Extrapolation Formulae}
\label{sec:ExtForm}
In addition to the effective Lagrangians given in Eqs.~(\ref{eq:Lfree1}-\ref{eq:Lnonlin}), we need some more ingredients. For our purposes, the most important one is probably the chiral Lagrangian which yields the leading quark mass insertions for the self-energies (see also \cite{Jenkins:1995vb,Urech:1995ry,Bijnens:1997ni,Bijnens:1997rv}), 
\begin{align}\label{eq:Lchi}
\mathcal{L}_{\chi}^{(0)} = b_{0}^{V}\langle V_{\mu}V^{\mu}\rangle\langle\chi_{+}\rangle + b_{D}^{V}\langle V_{\mu}\lbrace\chi_{+},V^{\mu}\rbrace\rangle + b_{0}^{VS}S_{\mu}S^{\mu}\langle\chi_{+}\rangle + b_{08}S_{\mu}\langle V^{\mu}\chi_{+}\rangle.
\end{align}
These $\mathcal{O}(p^2)$ contact terms including the octet-singlet mixing terms result in the following contributions to the self energies:
\begin{align}
\Pi_{\rho,\text{ct}} &= 8B_{0}\left(b_{0}^{V}(2m_{\ell}+m_{s})+2b_{D}^{V}m_{\ell}\right),\label{eq:treerho}\\
\Pi_{K^{*},\text{ct}} &= 8B_{0}\left(b_{0}^{V}(2m_{\ell}+m_{s})+b_{D}^{V}(m_{\ell}+m_{s})\right),\\
\Pi_{\phi^{(8)},\text{ct}} &= 8B_{0}\left(b_{0}^{V}(2m_{\ell}+m_{s})+\frac{2}{3}b_{D}^{V}(m_{\ell}+2m_{s})\right),\\
\Pi_{\phi^{(0)},\text{ct}} &= 8B_{0}b_{0}^{VS}\left(2m_{\ell}+m_{s}\right),\label{eq:treesing}\\
\Pi_{08,\text{ct}} &= 4B_{0}b_{08}\sqrt{\frac{2}{3}}(m_{\ell}-m_{s})=\Pi_{80,\text{ct}}.\label{eq:treemix}
\end{align}
\begin{figure}
\includegraphics[width=0.2\textwidth]{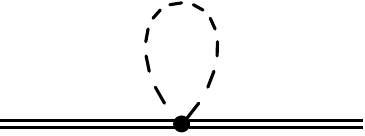}\hspace{1cm}\includegraphics[width=0.2\textwidth]{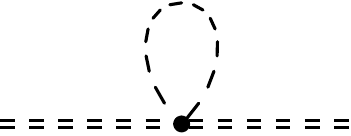}\hspace{1cm}\includegraphics[width=0.2\textwidth]{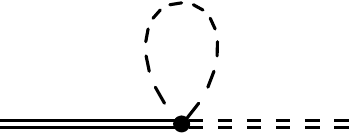}
\caption{The three tadpole diagrams we include in our calculation. The double line represents the octet vector mesons and the double dashed line represents the singlet vector meson. The dashed line stands for any of the pseudo goldstone bosons $\pi$, $K$ or $\eta$.}
\label{fig:tad}
\end{figure}
Note that the mixing term disappears as soon as we go to the $SU(3)$ symmetric limit $m_{\ell}=m_s$. The terms of Eq.~(\ref{eq:Lchi}) also lead to tadpole graphs (see Fig. \ref{fig:tad}) which also contain an octet-singlet mixing contribution proportional to $b_{08}$:
\begin{align}
\begin{split}
\Pi_{\rho,\text{tad}} &= -\frac{4B_{0}b_{0}^{V}}{F_{0}^{2}}\left(6m_{\ell}I_{\pi}+4(m_{\ell}+m_{s})I_{K}+\frac{2}{3}(m_{\ell}+2m_{s})I_{\eta}\right)\\
 &\quad- \frac{4B_{0}b_{D}^{V}}{F_{0}^{2}}\left(6m_{\ell}I_{\pi}+2(m_{\ell}+m_{s})I_{K}+\frac{2}{3}m_{\ell}I_{\eta}\right), 
\end{split}\\
\begin{split}
\Pi_{K^{\star},\text{tad}} &= -\frac{4B_{0}b_{0}^{V}}{F_{0}^{2}}\left(6m_{\ell}I_{\pi}+4(m_{\ell}+m_{s})I_{K}+\frac{2}{3}(m_{\ell}+2m_{s})I_{\eta}\right)\\
 &\quad- \frac{4B_{0}b_{D}^{V}}{F_{0}^{2}}\left(3m_{\ell}I_{\pi}+3(m_{\ell}+m_{s})I_{K}+\frac{1}{3}(m_{\ell}+4m_{s})I_{\eta}\right), 
\end{split}\\
\begin{split}
\Pi_{\phi^{(8)},\text{tad}} &= -\frac{4B_{0}b_{0}^{V}}{F_{0}^{2}}\left(6m_{\ell}I_{\pi}+4(m_{\ell}+m_{s})I_{K}+\frac{2}{3}(m_{\ell}+2m_{s})I_{\eta}\right)\\
 &\quad- \frac{4B_{0}b_{D}^{V}}{F_{0}^{2}}\left(2m_{\ell}I_{\pi}+\frac{10}{3}(m_{\ell}+m_{s})I_{K}+\frac{2}{9}(m_{\ell}+8m_{s})I_{\eta}\right),
\end{split}\\
\Pi_{\phi^{(0)},\text{tad}} &= -\frac{4B_{0}b_{0}^{VS}}{F_{0}^{2}}\left(6m_{\ell}I_{\pi}+4(m_{\ell}+m_{s})I_{K}+\frac{2}{3}(m_{\ell}+2m_{s})I_{\eta}\right),\\
\Pi_{08,\text{tad}} &= \Pi_{80}|_{tad} = -\frac{B_{0}b_{08}}{F_{0}^{2}}\sqrt{\frac{2}{3}}\left(6m_{\ell}I_{\pi}-2(m_{\ell}+m_{s})I_{K}+\frac{2}{3}(m_{\ell}-4m_{s})I_{\eta}\right),\label{eq:tadmix}
\end{align}
where the loop functions $I_{\pi,K,\eta}$ are defined in appendix \ref{app:Loops}. Again, the mixing contribution vanishes in the $SU(3)$ limit as it should. 
Of course, there are many more possible terms, with undetermined coefficients, which could generate tadpole graphs (see e.~g. ref.~\cite{Bijnens:1997ni}, which uses large-$N_{c}$ arguments to limit and constrain the corresponding parameters). So, strictly speaking, our calculation will only be complete at leading one-loop order $\mathcal{O}(p^3)$. We take along the tadpole results above only to be able to estimate such higher order effects later.
\begin{figure}
\centering
\subfigure[]{\includegraphics[width=0.2\textwidth]{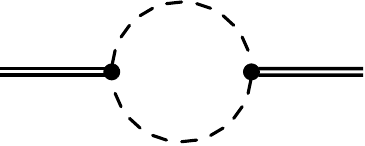}\label{fig:bubble}}\hspace{1cm}\subfigure[]{\includegraphics[width=0.2\textwidth]{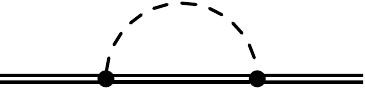}\label{fig:sunsetoct}}\hspace{1cm}\subfigure[]{\includegraphics[width=0.2\textwidth]{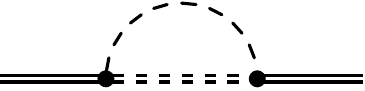}\label{fig:sunsetsing}}\\
\subfigure[]{\includegraphics[width=0.2\textwidth]{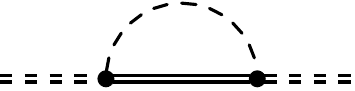}\hspace{1cm}\label{fig:sunsetsing2}}\subfigure[]{\includegraphics[width=0.2\textwidth]{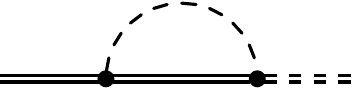}\label{fig:sunsetmix1}}
\caption{The other loop diagrams we include in our calculation. The double line represents the octet vector mesons and the double dashed line represents the singlet vector meson. The dashed line stands for any of the pseudo goldstone bosons $\pi$, $K$ or $\eta$.}
\label{fig:loop}
\end{figure}
All other one-loop contributions to the vector meson self-energies are shown in Fig. \ref{fig:loop}. 
Let us first investigate the bubble-type diagram shown in \ref{fig:bubble}, which only occurs for the octet vector mesons (see Eq.~(\ref{eq:Llin})). This diagram was not included in the 'heavy vector meson' approach \cite{Bijnens:1997ni}. The contribution due to the intermediate states with two PGBs is also absent in 'quenched' QCD \cite{Leinweber:1993yw}. Calculating this diagram results in both a contribution to the masses of the vector mesons as well as to the widths. The results for the bubble self-energy contributions to $\Pi^{PGB}_{T,loop}(s)$ for the respective octet members read
\begin{align}
\Pi_{\rho,\text{bbl}}(s) &= -\frac{g_{V}^{2}s^{2}}{F_{0}^{4}}\left(4I_{A}^{\pi\pi}(s) +2I_{A}^{\bar{K}K}(s)\right)  , \label{eq:PiRhoBbl}\\
\Pi_{K^{\star},\text{bbl}}(s) &= -\frac{g_{V}^{2}s^{2}}{F_{0}^{4}}\left(3I_{A}^{\pi K}(s) +3I_{A}^{K\eta}(s)\right)  ,\label{eq:PiKstarBbl}\\
\Pi_{\phi^{(8)},\text{bbl}}(s) &= -\frac{g_{V}^{2}s^{2}}{F_{0}^{4}}\left(6I_{A}^{\bar{K}K}(s)\right) ,\label{eq:PiPhi8Bbl}
\end{align}
The explicit mass corrections arising from the sunset-type diagrams figs. \ref{fig:sunsetoct}, \ref{fig:sunsetsing} and \ref{fig:sunsetsing2} take the following form:
\begin{align}
\Pi_{\rho,\text{sun}}(s) &= -\frac{4(g_{A}^{V})^{2}s}{F_{0}^{2}}\left(\frac{2}{3}I_{A}^{\pi V}(s) + 2I_{A}^{K\,V}(s) + \frac{2}{3}I_{A}^{\eta V}(s)\right)  -\frac{4(g_{A}^{VS})^{2}s}{F_{0}^{2}}I_{A}^{\pi S}(s), \\
\Pi_{K^{\star},\text{sun}}(s) &= -\frac{4(g_{A}^{V})^{2}s}{F_{0}^{2}}\left(\frac{3}{2}I_{A}^{\pi V}(s) + \frac{5}{3}I_{A}^{K\,V}(s) + \frac{1}{6}I_{A}^{\eta V}(s)\right)  -\frac{4(g_{A}^{VS})^{2}s}{F_{0}^{2}}I_{A}^{K\,S}(s),\\
\Pi_{\phi^{(8)},\text{sun}}(s) &= -\frac{4(g_{A}^{V})^{2}s}{F_{0}^{2}}\left(2I_{A}^{\pi V}(s) + \frac{2}{3}I_{A}^{K\,V}(s) + \frac{2}{3}I_{A}^{\eta V}(s)\right)  -\frac{4(g_{A}^{VS})^{2}s}{F_{0}^{2}}I_{A}^{\eta S}(s),\\
\Pi_{\phi^{(0)},\text{sun}}(s) &= -\frac{4(g_{A}^{VS})^{2}s}{F_{0}^{2}}\left(3I_{A}^{\pi V}(s) + 4I_{A}^{K\,V}(s) + I_{A}^{\eta V}(s)\right).
\end{align}
The last two diagrams shown in Fig. \ref{fig:loop} contribute to the singlet-octet mixing and take the form
\begin{align}
\Pi_{08,\text{sun}}(s) = -\frac{4g_{A}^{V}g_{A}^{VS}}{F_{0}^{2}}s\left(\sqrt{6}I_{A}^{\pi V}(s) - 2\sqrt{\frac{2}{3}}I_{A}^{K\,V}(s) - \sqrt{\frac{2}{3}}I_{A}^{\eta V}(s)\right).\label{eq:sunmix}
\end{align}
The above results for the contributions to $\Pi^{PGB}_{T,loop}(s)$ can be directly plugged in Eqs.~(\ref{eq:newselfenergy},\ref{eq:08matrix}), with $s_{V,pole}^{\star}=(M_{V}^{\star})^{2}=(855\,\mathrm{MeV})^{2}$ \cite{Bietenholz:2011qq}, and $s_{V,pole}^{\star}=(M_{S}^{\star})^{2}$, where $M_{S}^{\star}$ is an unknown parameter to be determined from the fits. The complex zeroes of Eq.~(\ref{eq:newselfenergy}) give the mass and the width of the corresponding vector meson, e.~g. $s_{\rho,pole}=M_{\rho}^{2}-iM_{\rho}\Gamma_{\rho}$ in the case of the $\rho$ meson.\\Having collected all the expressions for the loop contributions, we now have to discuss how we treat the loop integrals, with respect to regularization and power-counting. Here, it is important to realize that the loop integrals are in principle determined, up to some polynomials in $s$, by their corresponding threshold singularities and branch cuts, by means of a dispersive representation, see e.g. app.~\ref{app:Disp} for a demonstration, and Sect.~5 of \cite{Klingl:1996by} (dispersive representations of the $\pi\pi$ loop were also used in \cite{Chiang:1973qh,Benayoun:2000ti,Leinweber:1993yw}) . We evaluate all loop integrals employing dimensional regularization and use the $\overline{\text{MS}}$ scheme to deal with the ultraviolet divergences. Since a constant part of the loop contribution has been effectively absorbed in $(M_{V}^{\star})^{2}$, in the form of the subtraction $\Pi_{T,loop}^{\star PGB}(s_{pole}^{\star})$, and due to renormalization conditions like Eq.~(\ref{eq:rescond}), our renormalized loop corrections formally start at second chiral order, with terms of $\mathcal{O}((s-(M_{V}^{\star})^{2})^{2})$ and $\mathcal{O}(\delta m_{\ell})$. While our results for the loop portion of the self-energies $\Pi(s)$ can directly be mapped onto a dispersive representation, the power counting for the loop graphs is not straightforward, as already mentioned in the Introduction, and discussed in \cite{Bruns:2004tj,Djukanovic:2009zn}. It was demonstrated in \cite{Bruns:2004tj} that the genuine ``soft-pion'' part of the bubble diagram scales with the fractional power $M_{\phi}^{d}$ in dimensional regularization, which leads to an $\mathcal{O}(p^{4})$ contribution in $d\rightarrow 4$ space-time dimensions, and does not include the decay-threshold singularity, which could however be important phenomenologically. Also, from a na\"{i}ve power-counting, the sunset graphs should scale as $\mathcal{O}(p^{3})$ for $d\rightarrow 4$. Of course, one could in principle employ a chiral expansion of the loop graphs and absorb the real part of the $\mathcal{O}(p^{2})$ terms in the available counterterms. For the present application, however, a chiral expansion of the loop graphs is not effective due to the nearby presence of the $\varphi\varphi,V\varphi$ and $S\varphi$ decay thresholds. Therefore, for the purpose of the present application, we simply stick to the $\overline{\text{MS}}$ scheme (similar to the treatment of the nucleon self-energy in \cite{Gasser:1987rb}) in combination with dispersion-theoretic arguments, but note that one should keep all these subtleties in mind if one attempts a higher-order calculation within the present framework.\\To complete our collection of formulae, we also give the form of the counterterms $f_{1}$ occuring in Eq.~(\ref{eq:newselfenergy}), up to terms linear in the symmetry breaking $\delta m_{\ell}$:
\begin{align}
f_{1}^{\rho} &= f_{1}^{V\star}-8B_{0}z_{D}^{V}\delta m_{\ell} + \mathcal{O}((\delta m_{\ell})^{2}), \label{eq:f1Rho}\\
f_{1}^{K^{\star}} &= f_{1}^{V\star}+4B_{0}z_{D}^{V}\delta m_{\ell} + \mathcal{O}((\delta m_{\ell})^{2}),\label{eq:f1Kstar}\\
f_{1}^{\phi^{(8)}} &= f_{1}^{V\star}+8B_{0}z_{D}^{V}\delta m_{\ell} + \mathcal{O}((\delta m_{\ell})^{2}),\label{eq:f1Phi8}\\
f_{1}^{\phi^{(0)}} &= f_{1}^{S\star} + \mathcal{O}((\delta m_{\ell})^{2}),\label{eq:f1Phi0}\\
f_{1}^{08} &= 0 - 2\sqrt{6}B_{0}z_{08}\delta m_{\ell} + \mathcal{O}((\delta m_{\ell})^{2}).\label{eq:f108}
\end{align}
Counterterms of $\mathcal{O}((s-s_{pole}^{\star})^{2})$ and $\mathcal{O}((s-s_{pole}^{\star})(\delta m_{\ell})^{2})$ are neglected in the following, which sets the limits to our accuracy in the determination of the energy-dependence of the self-energies in the vector resonance region. 
To further clarify the origin of the above terms, note that e.g. the counterterms contributing to $\Pi_{08}$ could be derived from the following terms in a Lagrangian,
\begin{align}\label{eq:Lmix}
\mathcal{L}_{mix} =  b_{08}'S_{\mu}\langle V^{\mu}\chi_{+}\rangle - \frac{z_{08}}{4}S_{\mu\nu}\langle V^{\mu\nu}\chi_{+}\rangle + \ldots,
\end{align}
followed by a redefinition of the coupling (for a fixed numerical value of $M_V^{\star}$), $b_{08}'=b_{08}+z_{08}M_V^{\star 2}$. The introduction of additional terms (indicated by the dots in Eq.~(\ref{eq:Lmix})) would necessitate more complicated redefinitions, eventually leading to a polynomial in $(s-M_V^{\star 2})$ of higher degree. Similarly, the $z_{DV}$ terms above could be derived from a quark mass insertion like $-(z_{D}^{V}/4)\langle V_{\mu\nu}\lbrace\chi_{+},V^{\mu\nu}\rbrace\rangle$, and so on.
While $f_{1}^{V,S\star}$ and $z_{D}^{V}$ are determined from the condition $\mathrm{Re}R\overset{!}{=}1$, there is no natural way to fix $z_{08}$, so it should in principle be treated as a free parameter, in order to avoid any prejudice in the description of the singlet-octet mixing amplitude. 
\section{Results and discussion}
\label{sec:Results}
Before giving the results of our present work, we have to specify our numerical input and the data set we use to fit the undetermined parameters. Let us first discuss the decay constants of the pseudoscalar mesons (PGBs). Since we attempt an expansion around the reference point $\star$ instead of the chiral limit, we should replace $F_{0}\rightarrow F_{\star}$ in the loop contributions to the vector meson self energies (where the difference amounts to a two-loop effect anyway). To one-loop order, one finds for $F_{\star}(\bar{m})$ \cite{Gasser:1984gg,Bruns:2012eh}:
\begin{align}\label{eq:Fstar}
F_{\star} = F_{0}\left(1+\frac{2B_{0}\bar{m}}{(4\pi F_{0})^{2}}\left(64\pi^{2}(3L_{4}+L_{5})-3\log\left(\frac{\sqrt{2B_{0}\bar{m}}}{\mu}\right)\right)\right)+\mathcal{O}(\bar{m}^{2}).
\end{align}
Numerically, for fixed $\bar{m}$, we set $F_{\star}$ to the central value found in \cite{Bruns:2012eh} for our selected reference point, and fix $F_{\star}=112\,\mathrm{MeV}$ from now on. The only exception is the analysis of the running of $M_{V}^{\star}$ with $\bar{m}$, where we take some higher-order effects along and insert the expression of Eq.~(\ref{eq:Fstar}), with the same parameters as we used in \cite{Bruns:2012eh}. We also choose to fix the renormalization scale to $\mu=770\,\mathrm{MeV}$. For the $\rho\rightarrow\pi\pi$ decay width, we find within our present approximations for the loop graphs
\begin{align}\label{eq:GammaRho}
\Gamma_{\rho}=\frac{g_{V}^{2}M_{\rho}^{2}}{48\pi F_{\star}^{4}}\sqrt{M_{\rho}^{2}-4M_{\pi}^{2}}^{3}.
\end{align}
We fix $g_{V}$ by requiring that Eq.~(\ref{eq:GammaRho}) reproduces the experimentally known value of $150\,\mathrm{MeV}$ at the physical point, which yields $g_{V}=0.125$. Usually, one inserts the {\em pion} decay constant $F_{\pi}=92.4\,\mathrm{MeV}$ in the formula for the $\rho\rightarrow\pi\pi$ decay width, which leads to the smaller value $g_{V}=0.085$. We will also use this second value in a further set of fits, to estimate the impact of higher-order effects on our results.\\The input parameters which are probably afflicted with the largest theoretical uncertainties (besides the parameter $z_{08}$) are the couplings $g_{A}^{V}$ and $g_{A}^{VS}$ which are responsible for the sunset graph contributions. Comparing our Lagrangian Eq.~(\ref{eq:Lnonlin}) to the heavy vector meson Lagrangian of \cite{Jenkins:1995vb}, we find the correspondences $g_{A}^{V}\sim g_{2}$ and $g_{A}^{VS}\sim g_{1}$, for which this paper seems to favor the prediction of the nonrelativistic chiral quark model, so that
\begin{align}\label{eq:gAVSCQM}
g_{A}^{V}\sim g_{2}^{\chi qm}=\frac{3}{4},\qquad g_{A}^{VS}\sim g_{1}^{\chi qm}=\frac{\sqrt{3}}{2}.
\end{align}
In \cite{Bijnens:1997ni}, large-$N_{c}$ arguments are used to neglect the combination $g'\sim g_{A}^{VS}-\frac{2}{\sqrt{3}}g_{A}^{V}$, and the estimate quoted above then leads to $g\sim\frac{1}{2}g_{A}^{V}=\frac{3}{8}=0.375$. This reference also cites some other estimates, which amount to somewhat smaller values, $g\approx 0.3$. Together with the assumption $g'\sim 0$, this would amount to $g_{A}^{V}\sim 0.6$ and $g_{A}^{VS}\sim 0.7$. In our fits, we will use various different sets for the two axial couplings to get a handle on the theoretical uncertainty. We already remark here that this uncertainty is much less influenced by the (smaller) uncertainty in the parameters $F_{\star}$ and $M_{V}^{\star}$, which we therefore choose to fix in all our fits. We also add that the sunset graphs yield by far the dominant corrections to the tree level results in most cases.\\
The framework outlined in the previous sections is particularly adapted to analyze the lattice data of the QCDSF collaboration presented in \cite{Bietenholz:2011qq}, where the data leading to the so-called fan plots is generated by varying the flavor symmetry breaking quark mass combination $\delta m_{\ell}$ while keeping the average quark mass $\bar{m}$ fixed to its physical value. To the accuracy needed here, it is adequate to fix the PGB mass in the  $\delta m_{\ell}\rightarrow 0$ limit,
\begin{align}\label{eq:Mstarnum}
2B_{0}\bar{m}+\mathcal{O}(\bar{m}^{2}\log\bar{m})=M_{\star}^{2}\approx (412\,\mathrm{MeV})^{2},
\end{align}
see \cite{Bruns:2012eh} for more details. In the latter reference, we have also introduced a convenient measure for the symmetry breaking, 
\begin{align}
\nu=\frac{M_{\pi}^{2}-X_{\pi}^{2}}{X_{\pi}^{2}}=\frac{2B_{0}\delta m_{\ell}}{M_{\star}^{2}} + \mathcal{O}(\bar{m}\delta m_{\ell},(\delta m_{\ell})^{2}),
\end{align}
where $X_{\pi}^{2}=\frac{1}{3}(2M_{K}^{2}+M_{\pi}^{2})$. The symmetric point $\star$ is then given by $\nu=0$ together with Eq.~(\ref{eq:Mstarnum}). At the physical point, we have $\nu\approx -0.885$. In the fan plots, the vector meson masses are normalized to the mass combination \cite{Bietenholz:2011qq}
\begin{align}\label{eq:XRho}
X_{\rho}=\frac{1}{3}\left(2M_{K^{\star}}+M_{\rho}\right)=M_{V}^{\star}+\mathcal{O}(\nu^{2}).
\end{align}
Besides the fan plot data for the $\rho$ and the $K^{\star}$, we will also use three data points for the dependence of $M_{V}^{\star}$ on $M_{\star}\sim\sqrt{2B_{0}\bar{m}}$, for $M_{\star}\approx 307$, $357$ and $413$ MeV, see again \cite{Bietenholz:2011qq}. A fourth data point at higher $M_{\star}$ is excluded from the fit because we limit our data to sets where the PGB masses are all $\lesssim\,500\,\mathrm{MeV}$, so that the application of a one-loop approximation in a chiral Lagrangian framework can be justified.\\The analysis of the dependence $M_{V}^{\star}(\bar{m})$ is used to determine the vector meson mass in the chiral limit, $\mathring{M}_{V}$ (which does not appear in the other observables, where we have eliminated it in favor of $M_{V}^{\star}$), and the LEC $b_{0}^{V}$, which is also absorbed in $M_{V}^{\star}$ in those other observables (up to some higher-order tadpoles). For the fan plots, the most important parameter is $b_{D}^{V}$. The singlet mass appears in loop corrections to both $M_{V}^{\star}$ and the mass ratios displayed in the fan plots, but is mainly determined from the condition that the singlet-octet mixing determinant of Eq.~(\ref{eq:08det}) has zeroes at $s=M_{\omega,\phi}^{2}-iM_{\omega,\phi}\Gamma_{\omega,\phi}$ (see \cite{Nakamura:2010zzi}) at the physical point $\nu=-0.885$, which is enforced by including its absolute value at these two pole positions in the $\chi^{2}$ function (actually, we disregard $\Gamma_{\omega}$ here, because it is mostly generated by a two-loop effect, where three pions occur in an intermediate state). The parameter $b_{08}$ is determined only from the zeroes of the determinant and has no direct influence on the $\rho$ and $K^{\star}$ masses. To determine the subleading $z_{08}$-term, it would be necessary to include more accurate information on the energy-dependence of the mixing amplitude. In the large-$N_ {c}$ limit, the vector mesons form a nonet, given in our matrix notation by $N_{\mu}=V_{\mu}+\frac{1}{\sqrt{3}}\mathds{1}S_{\mu}$, and terms with additional flavor traces are suppressed (see e.g. \cite{Jenkins:1995vb,Bijnens:1997ni,Bijnens:1997rv,Urech:1995ry}). Comparing with such a Lagrangian, this implies relations like
\begin{align}\label{eq:largeNestimates}
b_{0}^{V}\approx 0,\qquad b_{08}-\frac{4}{\sqrt{3}}b_{D}^{V}\approx 0,\qquad z_{08}-\frac{4}{\sqrt{3}}z_{D}^{V}\approx 0,
\end{align}
(in the sense of a suppression by inverse powers of $N_{c}$). Here, we will not rely on such estimates in general. Only the last relation for $z_{08}$ will be used in one set of fits (fits of type 'A') where we set $z_{08}\rightarrow\frac{4}{\sqrt{3}}z_{D}^{V}$. In a second set (fits of type 'B') we will neglect this energy-dependent correction, and set $z_{08}\rightarrow 0$. In the second case, the energy dependence of the mixing amplitude is entirely given by the loop graphs.\\
The fit results for various combinations of the input parameters $g_{A}^{V(S)}$ are displayed in Tab.~\ref{tab:wtadpolesA} and \ref{tab:wtadpolesB} (including the tadpole shown in Sect.~\ref{sec:ExtForm}) and Tab.~\ref{tab:wotadpolesA} and \ref{tab:wotadpolesB} (without tadpoles).
\begin{table}[ht]
\caption{Fit results (type A) including tadpole contributions, where $g_{V},g_A^V$ and $g_A^{VS}$ have been used as input.}
\begin{ruledtabular}
\begin{tabular}{c c c c c c c c c c}
 fit  & $g_{V}$ & $g_{A}^{V}$ & $g_{A}^{VS}$ & $\mathring{M}_{V}$ (GeV) & $b_{0}^{V}$  & $b_{D}^{V}$ & $M_{S}^{\star}$ (GeV) & $b_{08}$  & color \\
\hline
 1A & 0.125 & 3/4 & $\sqrt{3}/2$ & 0.631 & 0.056 & 0.022 & 1.011 & 0.218 & black  \\ 
\hline
 2A & 0.125 & 0.6 & 0.7 & 0.627 & 0.054 & 0.019 & 1.000 & 0.237 & orange  \\
\hline
 3A & 0.125 & 1/2 & 1/2 & 0.625 & 0.053 & 0.017 & 0.988 & 0.249 & blue  \\
\hline
 4A & 0.125 & 0 & 0 & 0.618 & 0.051 & 0.015 & 0.979 & 0.266  & red \\
\hline
 5A & 0.085 & 3/4 & $\sqrt{3}/2$ & 0.693 & 0.031& 0.065 & 0.958 & 0.231 & black \\
\hline
 6A & 0.085 & 0.6 & 0.7 & 0.683 & 0.027 & 0.063 & 0.949 & 0.245 & orange \\
\hline
 7A & 0.085 & 1/2 & 1/2 & 0.678 & 0.025 & 0.062 & 0.938 & 0.253 & blue \\
\hline
 8A & 0.085 & 0   & 0   & 0.667 & 0.020 & 0.062 & 0.930 & 0.266 & red \\
\end{tabular} 
\end{ruledtabular}
\label{tab:wtadpolesA}
\end{table}
\begin{table}[ht]
\caption{Fit results (type B) including tadpole contributions, where $g_{V},g_A^V$ and $g_A^{VS}$ have been used as input.}
\begin{ruledtabular}
\begin{tabular}{c c c c c c c c c c}
 fit  & $g_{V}$ & $g_{A}^{V}$ & $g_{A}^{VS}$ & $\mathring{M}_{V}$ (GeV) & $b_{0}^{V}$  & $b_{D}^{V}$ & $M_{S}^{\star}$ (GeV) & $b_{08}$  & color \\
\hline
 1B & 0.125 & 3/4 & $\sqrt{3}/2$ & 0.636 & 0.060 & 0.017 & 0.935 & 0.208 & black \\ 
\hline
 2B & 0.125 & 0.6 & 0.7 & 0.630 & 0.057 & 0.015 & 0.924 & 0.220 & orange  \\
\hline
 3B & 0.125 & 1/2 & 1/2 & 0.626 & 0.054 & 0.015 & 0.914 & 0.227 & blue  \\
\hline
 4B & 0.125 & 0 & 0 & 0.618 & 0.051 & 0.015 & 0.902 & $\pm 0.238$ & red \\
\hline
 5B & 0.085 & 3/4 & $\sqrt{3}/2$ & 0.696 & 0.032 & 0.062 & 0.924 & 0.220 & black \\
\hline
 6B & 0.085 & 0.6 & 0.7 & 0.685 & 0.028 & 0.062 & 0.914 & 0.230 & orange \\
\hline
 7B & 0.085 & 1/2 & 1/2 & 0.679 & 0.025 & 0.061 & 0.904 & 0.237 & blue \\
\hline
 8B & 0.085 & 0   & 0   & 0.667 & 0.020 & 0.062 & 0.893 & $\pm 0.246$ & red \\ 
\end{tabular} 
\end{ruledtabular}
\label{tab:wtadpolesB}
\end{table}
\begin{table}[ht]
\caption{Fit results (type A) without tadpole contributions, where $g_{V},g_A^V$ and $g_A^{VS}$ have been used as input.}
\begin{ruledtabular}
\begin{tabular}{c c c c c c c c c c}
 fit & $g_{V}$ & $g_{A}^{V}$ & $g_{A}^{VS}$ & $\mathring{M}_{V}$ (GeV) & $b_{0}^{V}$  & $b_{D}^{V}$ & $M_{S}^{\star}$ (GeV) & $b_{08}$   \\
\hline
 1A' & 0.125 & 3/4 & $\sqrt{3}/2$ & 0.630 & 0.076 & 0.025 & 1.010 &  0.247  \\ 
\hline
 2A' & 0.125 & 0.6 & 0.7 & 0.626 & 0.073 & 0.021 & 0.999 & 0.268  \\
\hline
 3A' & 0.125 & 1/2 & 1/2 & 0.623 & 0.072 & 0.019 & 0.988 & 0.281  \\
\hline
 4A' & 0.125 & 0 & 0 & 0.617 & 0.068 & 0.017 & 0.979 & 0.300  \\
\hline
 5A' & 0.085 & 3/4 & $\sqrt{3}/2$ & 0.691 & 0.048 & 0.073& 0.957 & 0.261 \\
\hline
 6A' & 0.085 & 0.6 & 0.7 & 0.682 & 0.043 & 0.072 & 0.947 & 0.276 \\
\hline
 7A' & 0.085 & 1/2 & 1/2 & 0.677 & 0.040 & 0.071 & 0.937 & 0.285 \\
\hline
 8A' & 0.085 & 0   & 0   & 0.666 & 0.034 & 0.070 & 0.929 & 0.300 \\
\end{tabular} 
\end{ruledtabular}
\label{tab:wotadpolesA}
\end{table}
\begin{table}[ht]
\caption{Fit results (type B) without tadpole contributions, where $g_{V},g_A^V$ and $g_A^{VS}$ have been used as input.}
\begin{ruledtabular}
\begin{tabular}{c c c c c c c c c c}
 fit & $g_{V}$ & $g_{A}^{V}$ & $g_{A}^{VS}$ & $\mathring{M}_{V}$ (GeV) & $b_{0}^{V}$  & $b_{D}^{V}$ & $M_{S}^{\star}$ (GeV) & $b_{08}$  \\
\hline
 1B' & 0.125 & 3/4 & $\sqrt{3}/2$ & 0.635 & 0.080 & 0.019 & 0.934 & 0.234  \\ 
\hline
 2B' & 0.125 & 0.6 & 0.7 & 0.629 & 0.076 & 0.018 & 0.923 & 0.247  \\
\hline
 3B' & 0.125 & 1/2 & 1/2 & 0.625 & 0.073 & 0.017 & 0.913 &  0.256 \\
\hline
 4B' & 0.125 & 0 & 0 & 0.617 & 0.068 & 0.017 & 0.901 & $\pm 0.268$ \\
\hline
 5B' & 0.085 & 3/4 & $\sqrt{3}/2$ & 0.694 & 0.050 & 0.071 & 0.923 & 0.247 \\
\hline
 6B' & 0.085 & 0.6 & 0.7 & 0.684 & 0.044 & 0.070 & 0.912 & 0.259 \\
\hline
 7B' & 0.085 & 1/2 & 1/2 & 0.678 & 0.040 & 0.070 & 0.902 & 0.266 \\
\hline
 8B' & 0.085 & 0   & 0   & 0.666 & 0.034 & 0.070 & 0.891 & $ \pm0.277$ \\
\end{tabular} 
\end{ruledtabular}
\label{tab:wotadpolesB}
\end{table}
\begin{figure}
\includegraphics[width=0.75\textwidth]{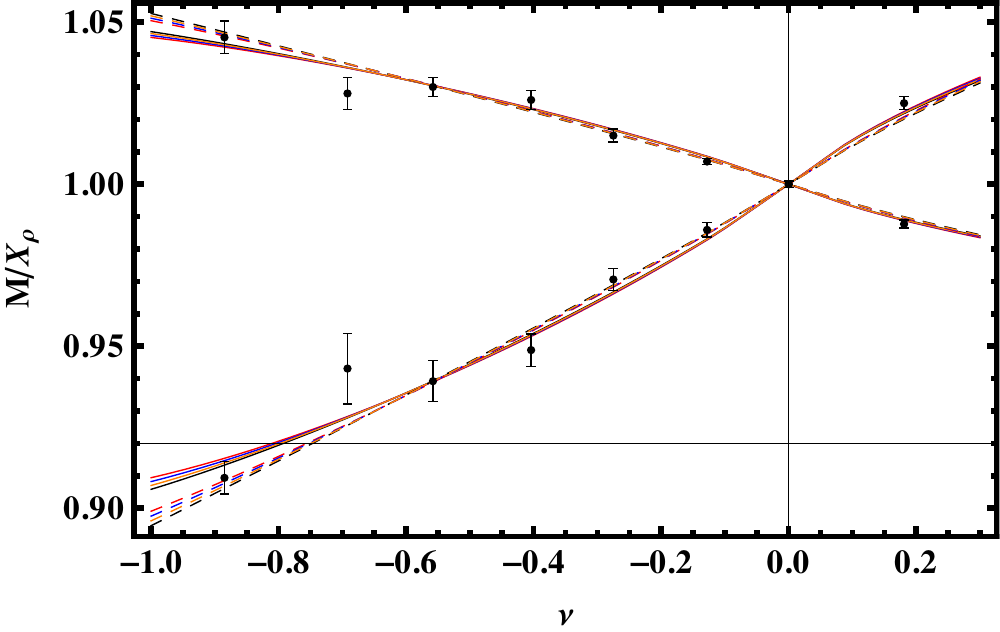}
\caption{The ratios $M_{\rho}/X_{\rho}$ and $M_{K^*}/X_{\rho}$ plotted for all parameter sets from Tab.~\ref{tab:wtadpolesA}. The color code for the different curves is shown in the tables above. Full lines: 1A-4A, dashed lines: 5A-8A. $X_{\rho}$ is defined in Eq.~(\ref{eq:XRho}).}
\label{fig:fanplots}
\end{figure}
\begin{figure}
\includegraphics[width=0.75\textwidth]{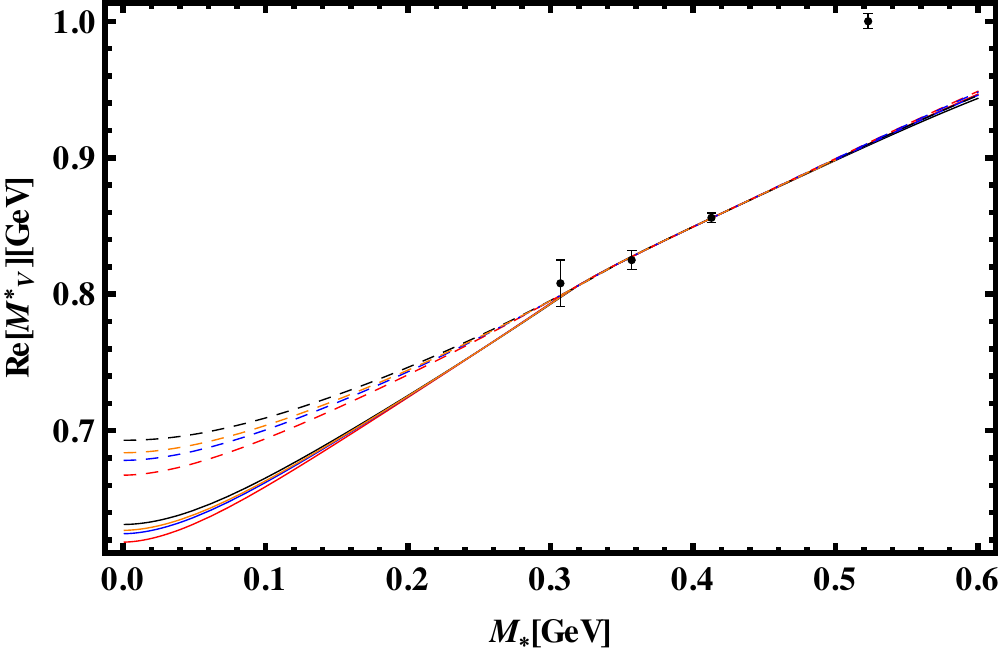}
\caption{The symmetric mass $M_V^{\star}$ (octet vector meson mass for $\delta m_{\ell}=0$) plotted for all parameter sets from Tab.~\ref{tab:wtadpolesA}. The color code for the different curves is shown in the tables above. Full lines: 1A-4A, dashed lines: 5A-8A.}
\label{fig:MVM}
\end{figure}
\begin{figure}
\subfigure[1A-4A, $\nu=-0.885$]{\includegraphics[width=0.37\textwidth]{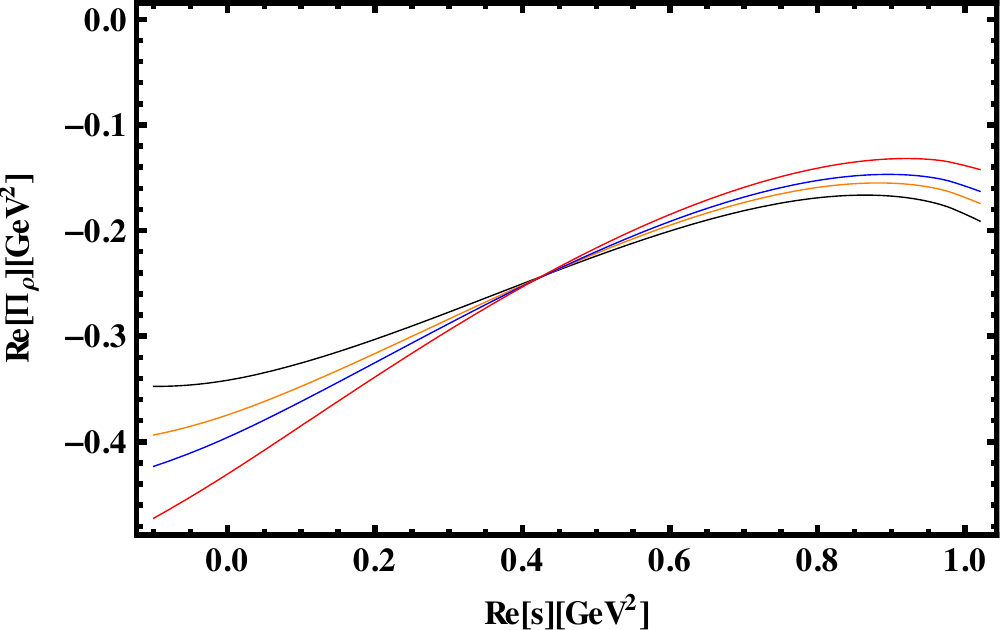}}\hspace{1cm}\subfigure[1A-4A, $\nu=0$]{\includegraphics[width=0.37\textwidth]{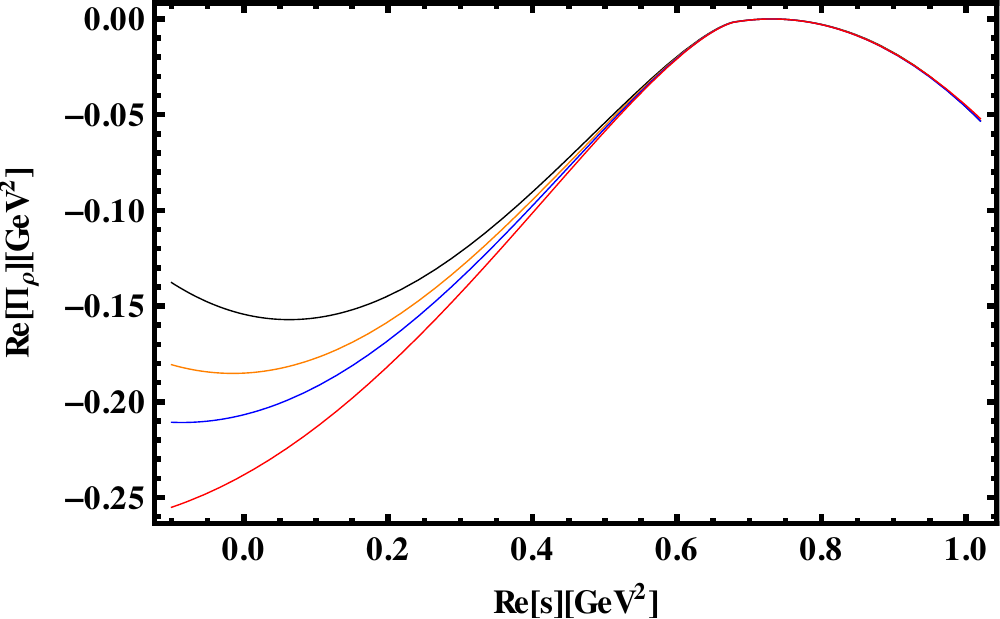}}\\
\caption{The energy dependence of the real part of $\Pi_{\rho}$ plotted for the parameter sets 1A-4A for $\nu=-0.885$ (left) and $\nu=0$ (right). The color code for the different curves is shown in the tables above.}
\label{fig:RePiRho}
\end{figure}
\begin{figure}
\subfigure[1A-4A, $\nu=-0.885$]{\includegraphics[width=0.37\textwidth]{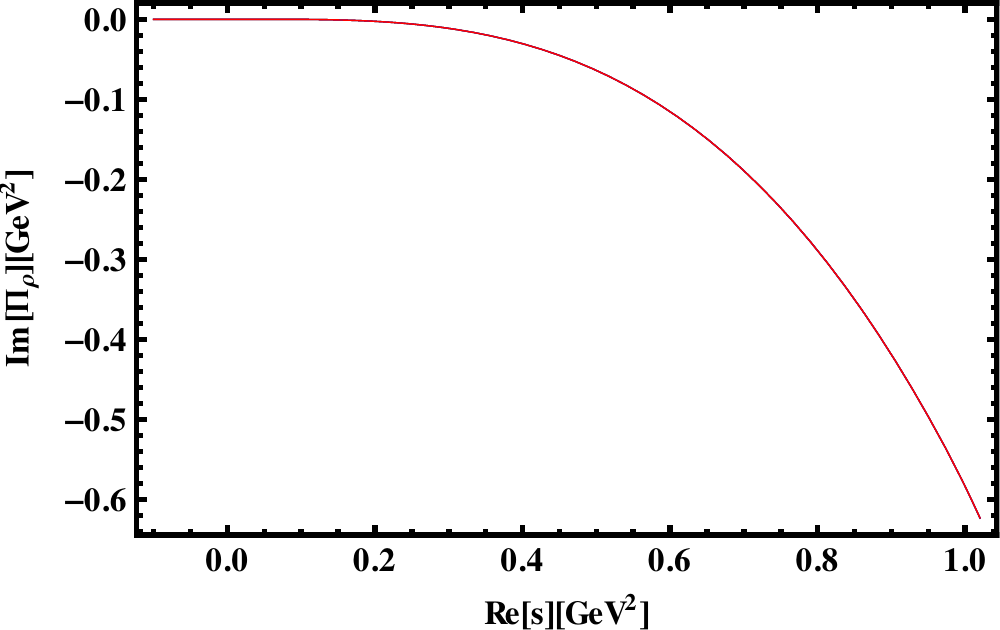}}\hspace{1cm}\subfigure[1A-4A, $\nu=0$]{\includegraphics[width=0.37\textwidth]{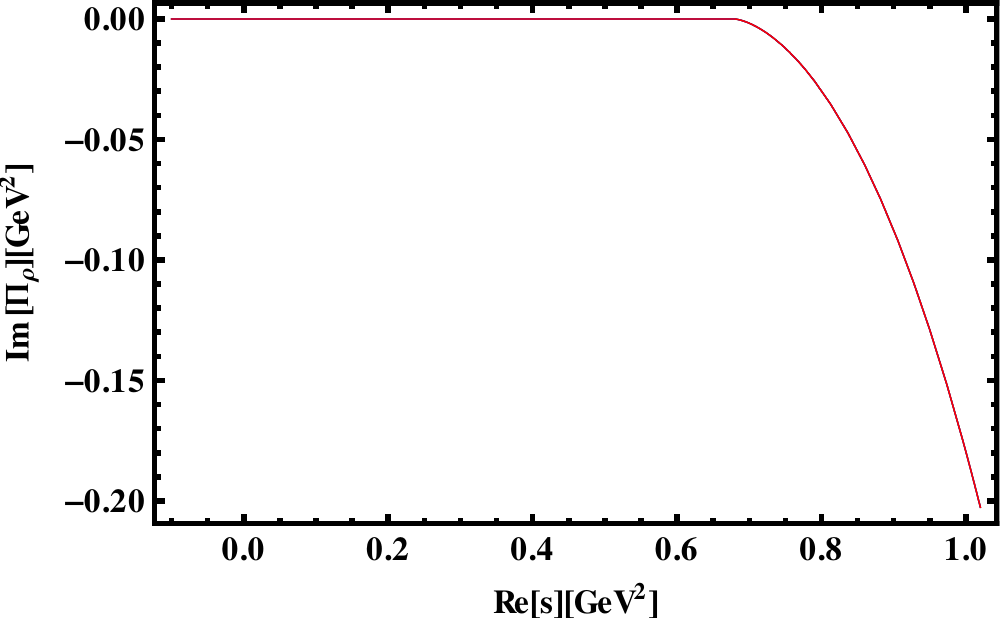}}\\
\caption{The energy dependence of the imaginary part of $\Pi_{\rho}$ plotted for the parameter sets 1A-4A for $\nu=-0.885$ (left) and $\nu=0$ (right). The color code for the different curves is shown in the tables above.}
\label{fig:ImPiRho}
\end{figure}
\begin{figure}
\subfigure[1A-4A]{\includegraphics[width=0.37\textwidth]{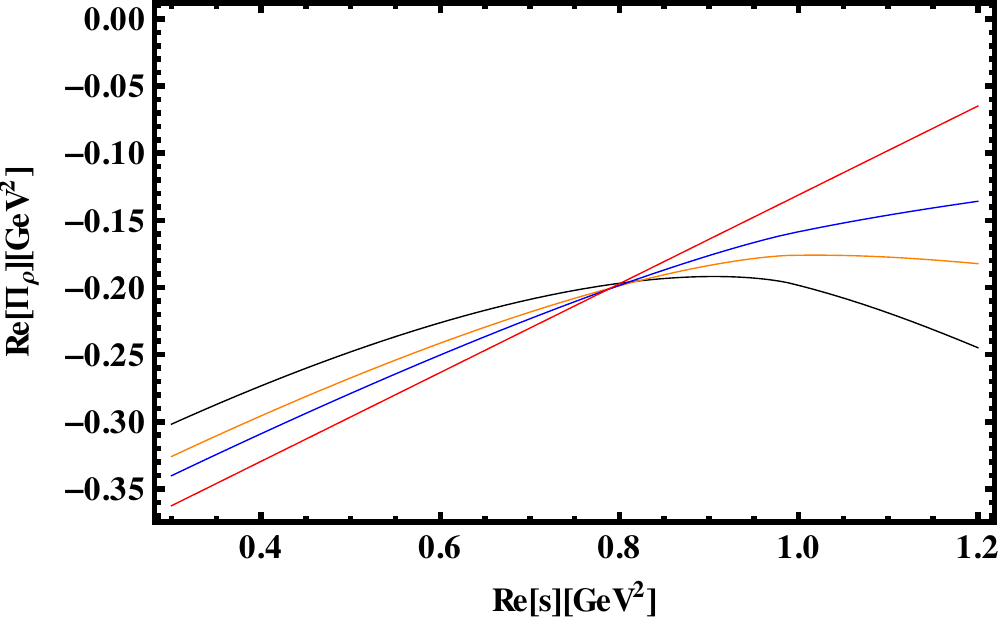}}\hspace{1cm}\subfigure[5A-8A]{\includegraphics[width=0.37\textwidth]{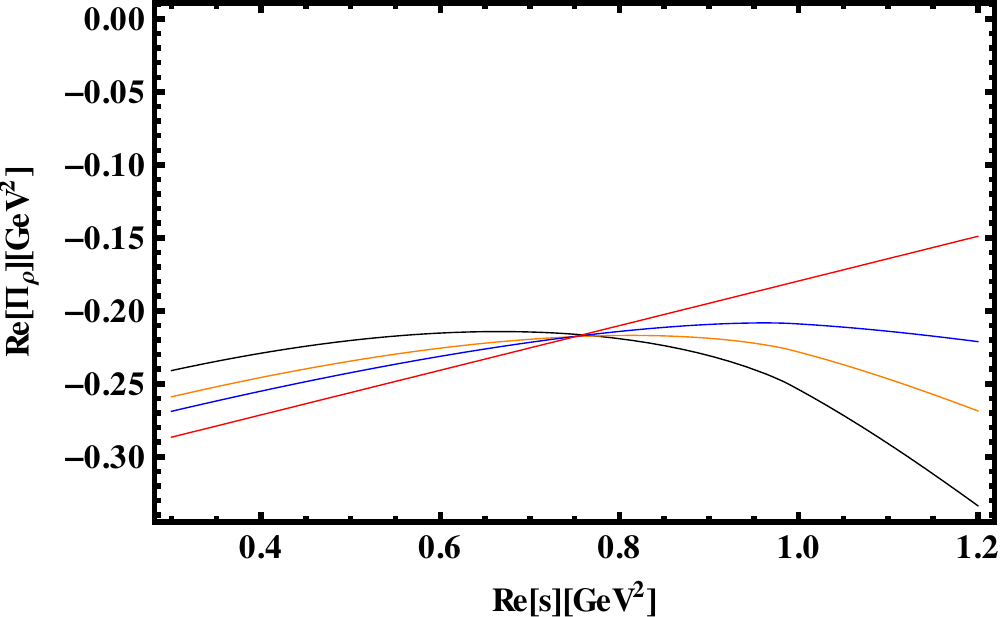}}\\
\subfigure[1B-4B]{\includegraphics[width=0.37\textwidth]{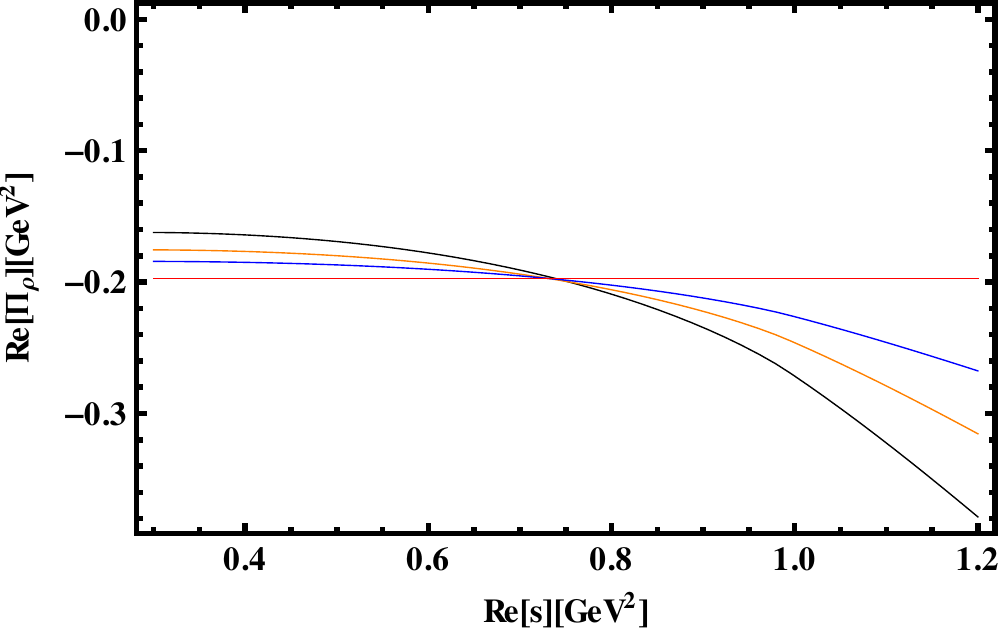}}\hspace{1cm}\subfigure[5B-8B]{\includegraphics[width=0.37\textwidth]{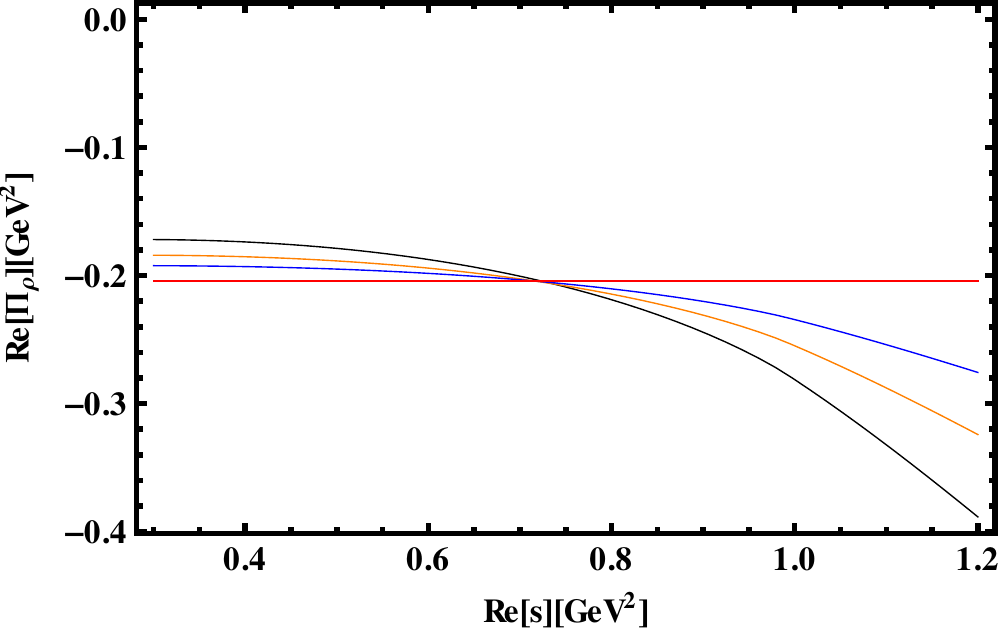}}\\
\caption{The energy dependence of $\Pi_{08}$ plotted for all parameter sets including tadpoles. The fit sets without tadpoles have been omitted due to their similarity. The color code for the different curves is shown in the tables above.}
\label{fig:Pi08}
\end{figure}
\begin{figure}
\subfigure[2A]{\includegraphics[width=0.37\textwidth]{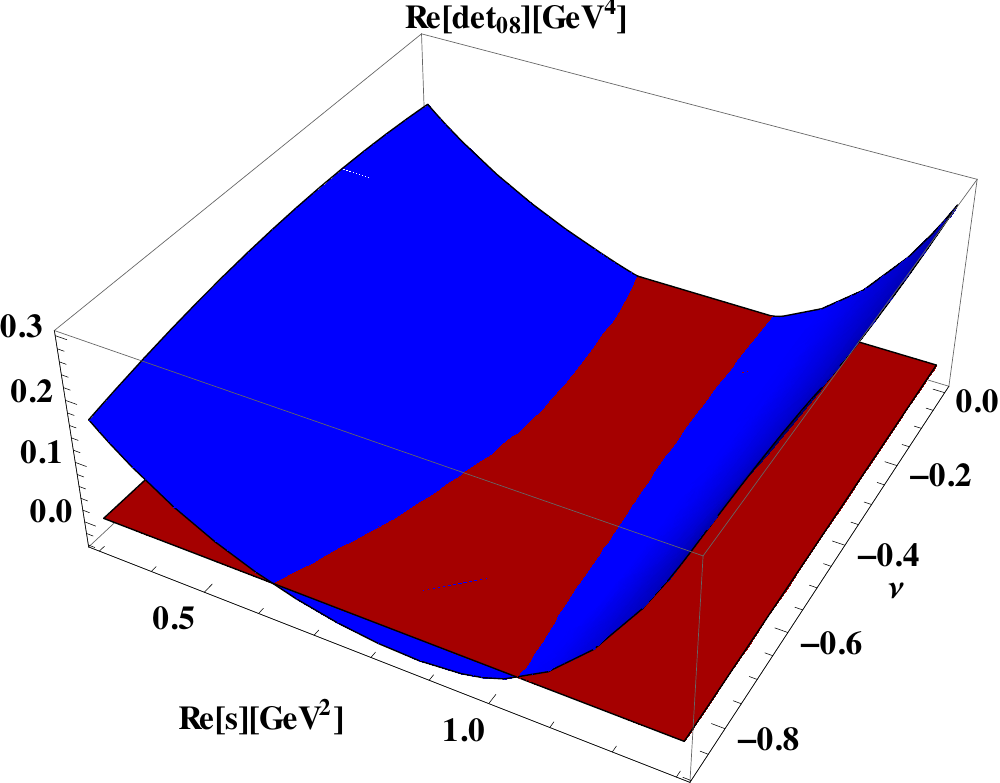}}\hspace{1cm}\subfigure[6A]{\includegraphics[width=0.37\textwidth]{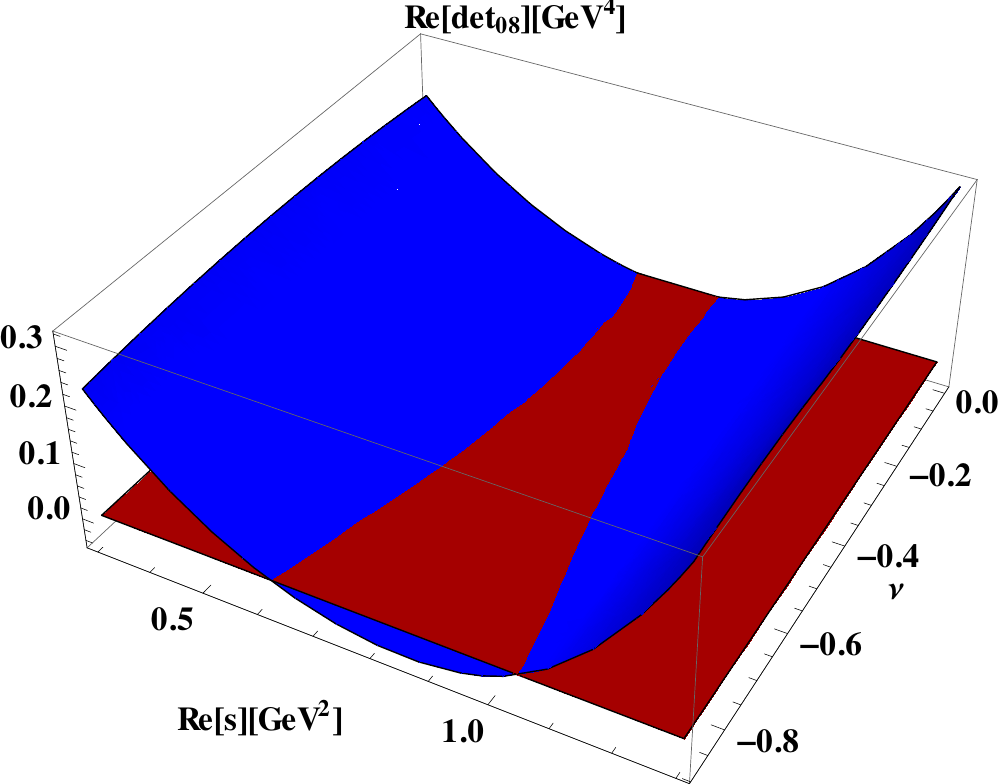}}\\
\subfigure[2B]{\includegraphics[width=0.37\textwidth]{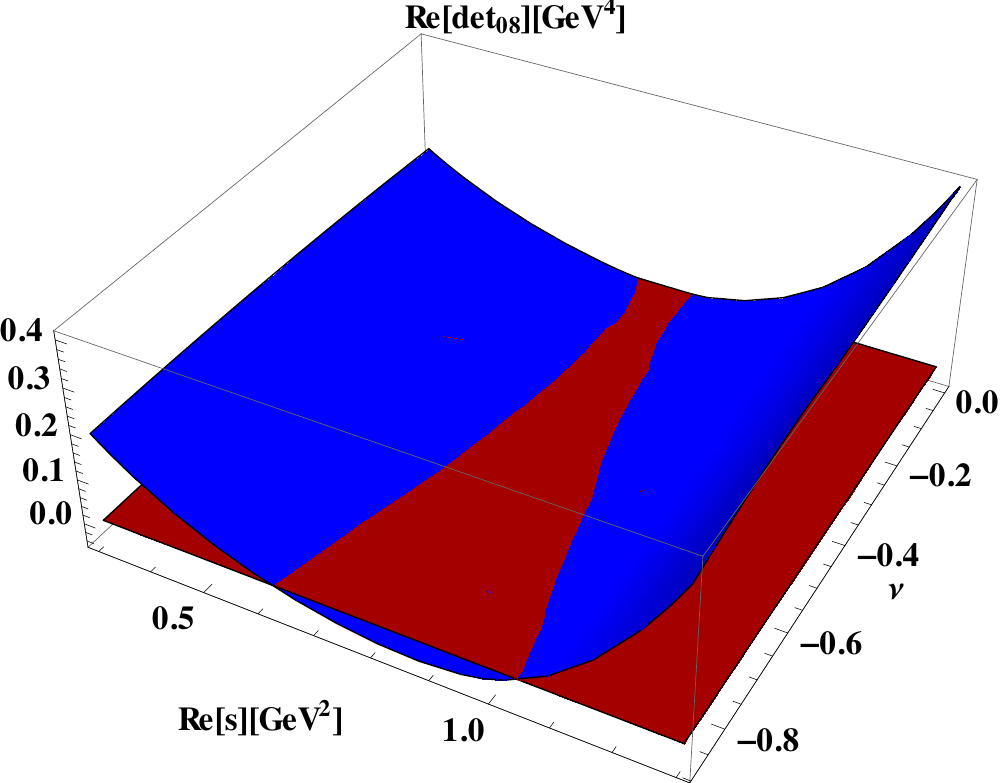}}\hspace{1cm}\subfigure[6B]{\includegraphics[width=0.37\textwidth]{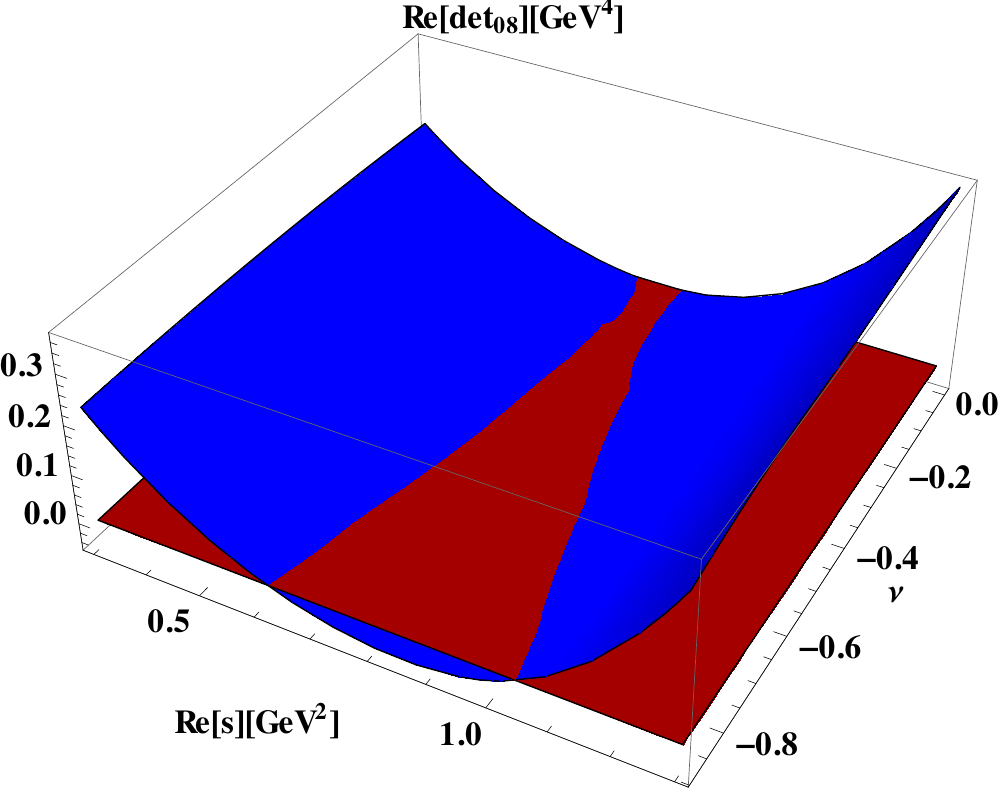}}\\
\caption{The real part of $\text{det}_{08}$ plotted in the $(\text{Re}[s],\nu)$ plane for fits 2A,2B,6A,6B.}
\label{fig:det083D}
\end{figure}
\begin{figure}
\subfigure[2A]{\includegraphics[width=0.37\textwidth]{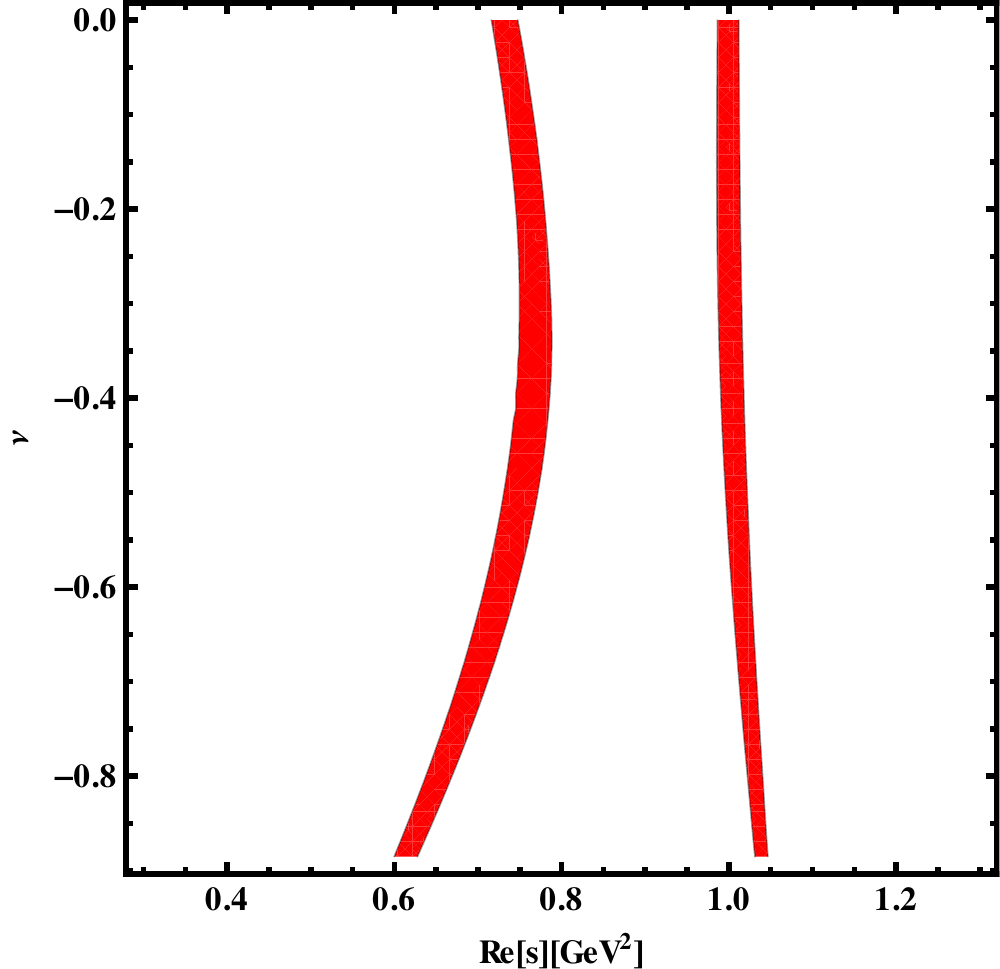}}\hspace{1cm}\subfigure[6A]{\includegraphics[width=0.37\textwidth]{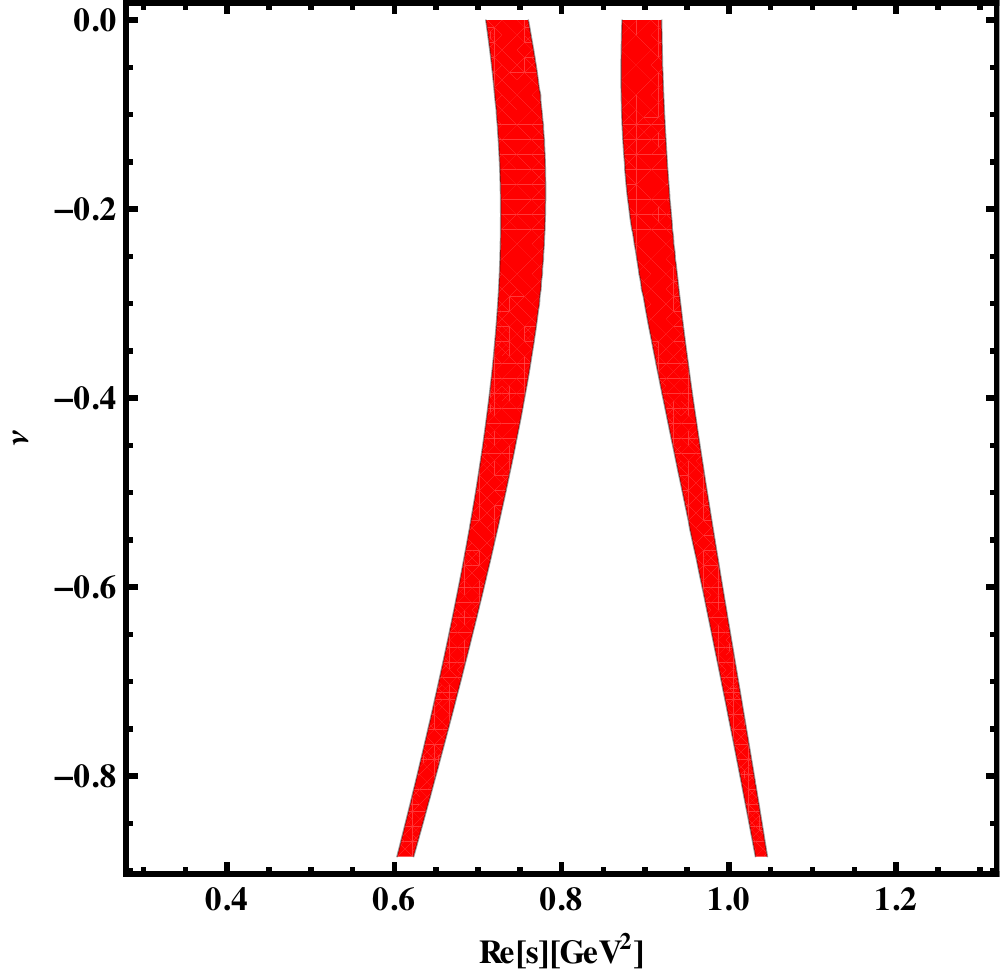}}\\
\subfigure[2B]{\includegraphics[width=0.37\textwidth]{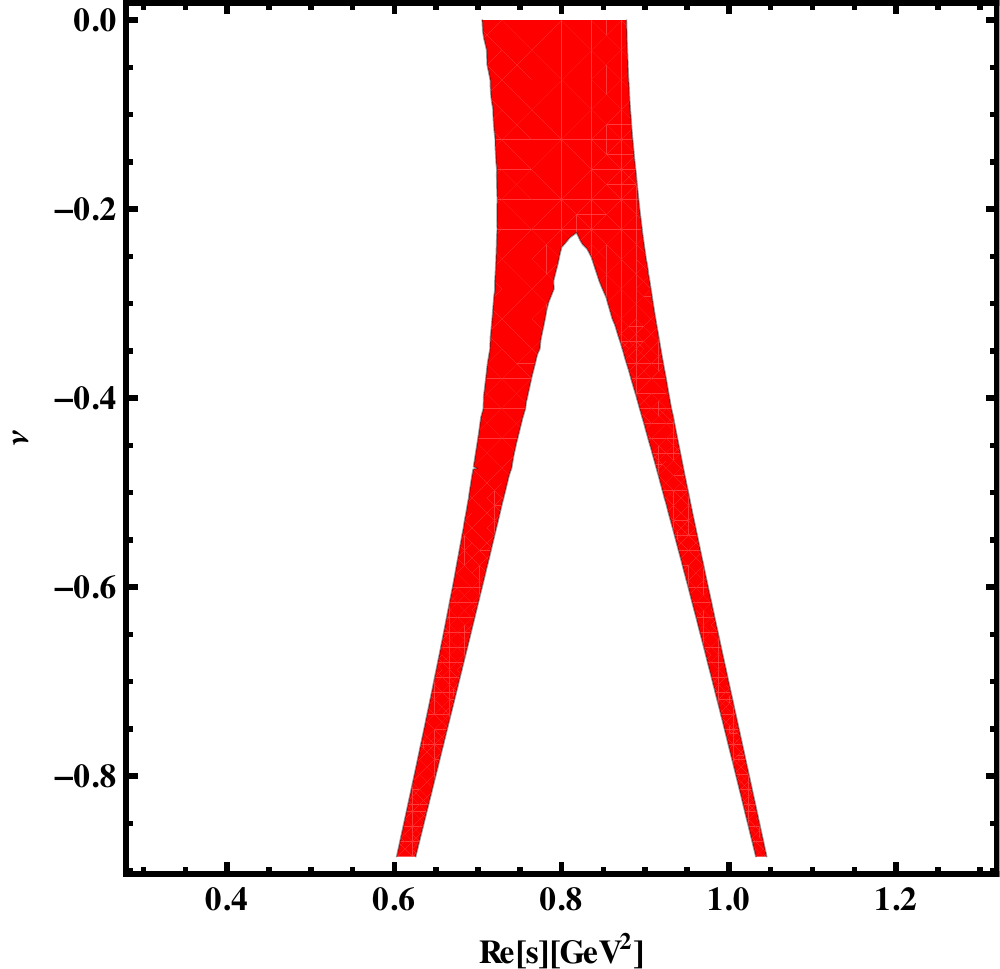}}\hspace{1cm}\subfigure[6B]{\includegraphics[width=0.37\textwidth]{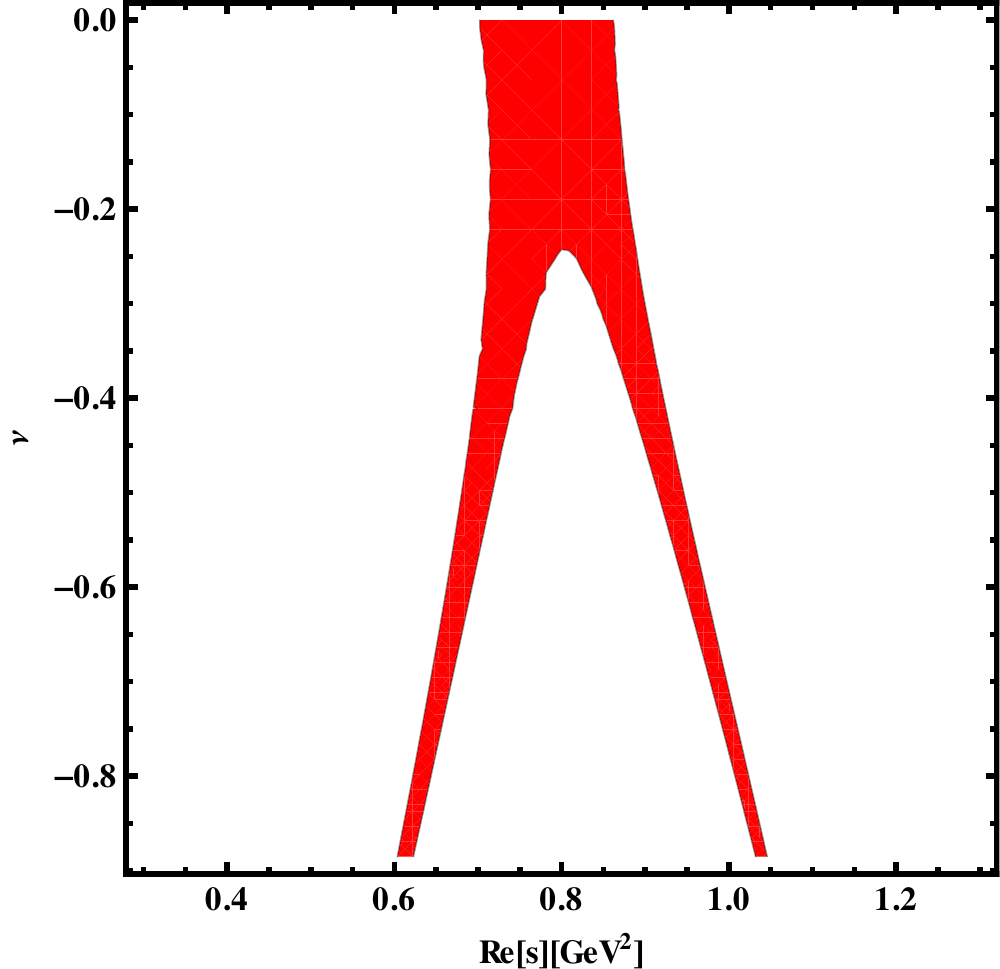}}\\
\caption{The plots depict the absolute value of $\text{Re}[\text{det}_{08}]$. The areas where $|\text{Re}[\text{det}_{08}]|<0.004\,\text{GeV}^4$ holds are colored red.}
\label{fig:contour}
\end{figure}
The fan plots for these sets of parameters are shown in Fig. \ref{fig:fanplots}. The lines resulting from the different fits can be barely distinguished. In the relative vicinity of the symmetric point at $\nu=0$, this is expected because our calculated corrections to $M_{V}^{\star}$ are of $\mathcal{O}(\delta m_{\ell})$. It seems that the variation of the input parameters can be almost completely compensated by the shift of the fitted parameters displayed in the tables above. Specifically, one observes that the LEC $b_{D}^{V}$ absorbs a large contribution from the real part of the bubble type loop graph $\sim g_{V}^{2}/F_{\star}^{4}$. In other words, the functional form of our leading one-loop expressions fixes the shape of the curves to a large extent, at least in the parameter range specified above. To a lesser extent this also applies for the dependence of $M_{V}^{\star}$ on $M_{\star}$ which is shown in Fig.~\ref{fig:MVM}. Here, however, the variation of $g_{V}$ has some effect on the determination of $\mathring{M}_{V}$, and obviously some more data points for lower $M_{\star}$ would be needed for a more accurate determination of the behavior of $M_{V}^{\star}(\bar{m})$ close to the chiral limit. One should be warned that the theoretical uncertainty is also larger in the region $M_{\star}\gtrsim\,400\mathrm{MeV}$, where one certainly should not trust a leading one-loop representation. This is also borne out by the fact that the fourth data point at higher $M_{\star}$, which was not included in the fits, is missed by the collection of these lines. As a side remark, we note that a very slight cusp due to the $V_{\star}\rightarrow\varphi_{\star}\varphi_{\star}$ decay threshold is visible in Fig.~\ref{fig:MVM} (at about $M_{\star}\sim 330\,\mathrm{MeV}$). The effect does not seem to be of big importance here - this would probably be different for scalar meson resonances \cite{Guo:2011gc}. \\ In Fig.~\ref{fig:RePiRho} and \ref{fig:ImPiRho}, we show the real and the imaginary part of the self-energy function for the $\rho$, for some typical fit results from Tab.~\ref{tab:wtadpolesA}. The fact that the curve for the real part for $\nu=0$ is tangent to the $s-$axis at $s=M_{V}^{\star 2}$ just reflects our chosen renormalization conditions. The energy dependence of $\Pi_{08}(s)$ is shown in Fig.~\ref{fig:Pi08}. Here the different sets can be well distinguished, but the trend is always the same: Our fits obviously favor a limited, but non-negligible energy dependence of the mixing amplitude, which can be partly traced back to the loop graphs, but also to the counterterm coefficient $z_{08}$ in the type 'A' fits. We note in passing that a strong energy dependence and a possible sign change of the mixing amplitude in the energy region between the $\omega$ and the $\phi$ mass has been observed in \cite{Bijnens:1997ni} (see their Table~2).\\Finally, the real part of the determinant $\mathrm{det}_{08}$ for four typical fits is shown in Fig.~\ref{fig:det083D} in the $(\text{Re}[s],\nu)$-plane, so that its energy dependence and its zeroes can be read off nicely for the range of the flavor symmetry breaking variable $\nu$ examined here.  The contour plots for the same fit sets of Tab.~\ref{tab:wtadpolesA} and \ref{tab:wtadpolesB} showing the positions of the zeroes of $\text{Re}[\text{det}_{08}]$ are displayed in Fig.~\ref{fig:contour}, to illustrate the running of the (real part of the) zeros of the determinant when tuning the symmetry-breaking variable $\nu$. \\In the results collected above, we have discarded a second class of fits where additional unphysical states appear in the singlet-octet sector, in the energy region where the present calculation should be applicable. This class preferredly results if the axial couplings $g_{A}^{V(S)}$ are large. We give an example for such an alternative scenario in the following Tab.~\ref{tab:sickfit}.
\begin{table}[h]
\caption{Result for a fit showing spurious states.}
\begin{ruledtabular}
\begin{tabular}{c c c c c c c c c}
 fit & $g_{V}$ & $g_{A}^{V}$ & $g_{A}^{VS}$ &  $\mathring{M}_{V}$ (GeV) & $b_{0}^{V}$  & $b_{D}^{V}$ & $M_{S}^{\star}$ (GeV) & $b_{08}$  \\
\hline
 $\widetilde{\mathrm{1A}}$ & 0.125 & 3/4 & $\sqrt{3}/2$ & 0.611 & 0.046 & 0.032 & 1.199 & 0.001  \\ 
\end{tabular}  
\end{ruledtabular}
\label{tab:sickfit}
\end{table}
Here, $b_{08}$ is much smaller than in the earlier fits (and $M_{S}^{\star}$ is notably larger), while the other parameters are in accord with the previous class of fits.
\begin{figure}
\subfigure[1A]{\includegraphics[width=0.45\textwidth]{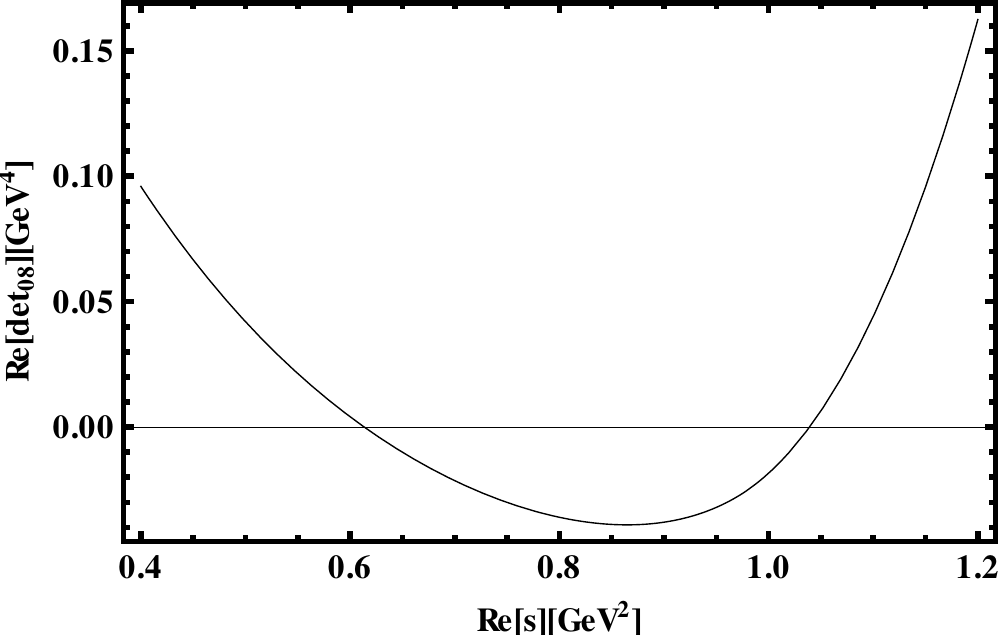}}\hspace{1cm}\subfigure[$\widetilde{\mathrm{1A}}$]{\includegraphics[width=0.45\textwidth]{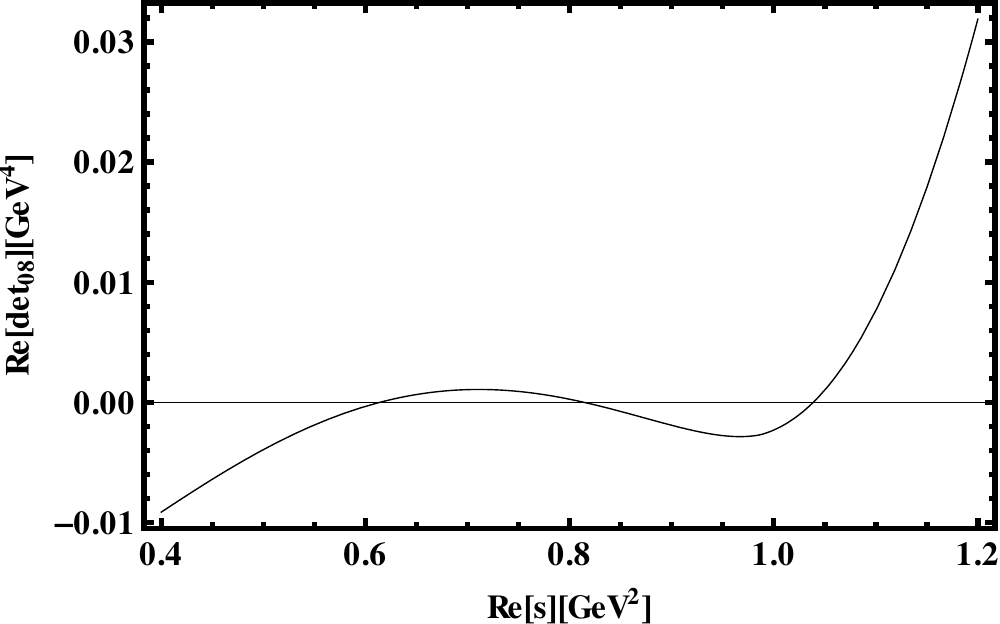}}
\caption{$\text{Re}[\text{det}_{08}]$ at $\nu=-0.885$ for a standard fit (1A of Tab.~\ref{tab:wtadpolesA}) (left) and for the fits of Tab.~\ref{tab:sickfit} showing spurious states (right)}
\label{fig:goodbaddet}
\end{figure}
In Fig.~\ref{fig:contour}, we plot the typical behavior of the determinant $\det_{08}$ for the two classes of fits. All the fits we have obtained could be grouped in one of the two classes, either the standard fits of Tab.~\ref{tab:wtadpolesA}-\ref{tab:wotadpolesB} or fits showing spurious states, with relatively large $M_{S}^{\star}$ and $g_{A}^{V(S)}$ but smaller $b_{08}$. Whereas the resulting plots for the masses look very much the same as in the 'standard' case, $\det_{08}$ shows a very different 'non-parabolic' behavior. The resulting parameter $z_{08}$ for fit $\widetilde{\mathrm{1A}}$ is $0.734\,\mathrm{GeV}^{-2}$, while it ranges between $0.826\ldots 0.900\,\mathrm{GeV}^{-2}$ for the fits 1A-4A and between $0.374\ldots 0.416\,\mathrm{GeV}^{-2}$ for 5A-8A.\\While the formulae for the vector meson masses and widths are accurate and model-independent to one-loop order, and to $\mathcal{O}(p^3)$ in the chiral counting, the question of the form of the energy-dependence of the two-point amplitudes is a subtle one and certainly deserves further study. In our opinion, our results at least show that an analysis that does not take into account the possibility of a sizeable variation of the mixing amplitude between $s=M_{\omega}^{2}$ and $s=M_{\phi}^{2}$ would not be under sufficient theoretical control. It would be very interesting to have lattice data for the variation of the $\omega$ and $\phi$ masses for different values of the symmetry breaking quark mass difference $\delta m_{\ell}\sim\nu$ in order to check whether the mixing scenario we have outlined here (see e.g. Fig.~\ref{fig:det083D}) is realistic. Our analysis of the mixing here could of course only be on a qualitative level. For a more accurate quantitative discussion of the dynamics in this sector, one has to consider vector meson decays, one has to take care of the corresponding relevant final-state interactions, and so on (see e.g. \cite{Klingl:1996by,Terschlusen:2013iqa,Lutz:2008km,Niecknig:2012sj,Schneider:2012ez}). Such an analysis is beyond the scope of the present investigation.\\As a general remark, we add that we do not observe a clear 'large-$N_{c}$' pattern in our results. The LEC $b_{0}^{V}$ is not notably suppressed compared with $b_{D}^{V}$, and the mass-splitting $M_{S}^{\star}-M_{V}^{\star}$ comes out small but non-negligible. Also, $b_{08}$ is somewhat larger than expected from the relation in Eq.~(\ref{eq:largeNestimates}) (independently of the scenario chosen to fix $z_{08}$). Of course, such a pattern can also not be ruled out by our findings, keeping in mind all the theoretical uncertainties discussed above. In particular, given the present data base, the result for $b_{0}^{V}$ is quite sensitive to higher-order effects, see e.g. tabs.~\ref{tab:wtadpolesA}-\ref{tab:wotadpolesB}. An extension of the present calculation to a next-to-leading one-loop calculation, without assuming a large-$N_{c}$ scaling, meets some difficulties: First, many additional tadpole graph contributions appear, and to fix the corresonding new LECs, one would need e.g. data on $V\varphi\rightarrow V\varphi$. Second, fan plot data at different values of the average quark mass $\bar{m}$ would be needed to fix all the quark mass insertions at the fourth chiral order. Moreover, at this order, one might also want to include electromagnetic contributions, and isospin breaking due to the light quark mass difference, which we have completely neglected here (of course, these effects are also not present in the lattice data used here). For numerical estimates of these effects, we again refer to \cite{Bijnens:1997ni} (see also \cite{Kucukarslan:2006wk} for the corresponding contributions to $\phi-\omega$ mixing). In a future study, finite-volume corrections should also be included, although these were claimed to cancel to a large extent in the mass ratios leading to the fan plots \cite{Bietenholz:2011qq}. In summary, the results of the present exploratory study, together with the results obtained on the extrapolation of baryon mass ratios in our previous work \cite{Bruns:2012eh}, lead us to conclude that the simulation strategy proposed and examined in \cite{Bietenholz:2011qq}, namely the extra\-polation from the 'symmetric point' (instead of the chiral limit) to physical quark masses is very promising, and that Chiral Perturbation Theory can be used as a reliable tool to guide such an extra\-polation, while the uncertainties for an expansion around the chiral limit are probably too large in the three-flavor case (see also the discussion in \cite{Bruns:2012eh} for the case of baryon masses, and references cited therein). In particular, for vector mesons we have the further advantage that these particles are (almost) stable particles at the symmetric point, so that the widths are given by symmetry-breaking corrections calculated perturbatively, and are dominated by two-particle channels in the vicinity of this reference point. Of course, additional data for smaller average quark masses would be welcome in order to determine the LECs more accurately, and to relate the results of the present application to the more common extrapolations to the chiral limit.
\appendix
\section{Dispersive representation of the $\mathbf{\pi\pi}$ loop}
\label{app:Disp}
Restoring the energy-dependence of the decay width of Eq.~(\ref{eq:GammaRho}),
\begin{align}
-iM_{\rho}\Gamma_{\rho}\rightarrow -i\frac{g_{V}^{2}}{6\pi F_{0}^{4}}(k^{2})^{\frac{3}{2}}|\vec{q}_{cms}^{\,\pi\pi}|^{3}=-i\frac{g_{V}^{2}}{48\pi F_{0}^{4}}(k^{2})^{3}\sqrt{1-\frac{4M_{\pi}^{2}}{k^{2}}}^{3},
\end{align}
where 
\begin{align}
|\vec{q}_{cms}^{\,\pi\pi}| = \frac{1}{2}\sqrt{k^{2}-4M_{\pi}^{2}},
\end{align}
we can insert the emerging expression in a dispersion relation with four subtractions ($k^{2}\equiv s$),
\begin{align}\label{eq:Pirhopipi}
\Pi_{\rho}^{\pi\pi}(s) =  c_{0}+c_{1}s+c_{2}s^{2}+c_{3}s^{3}-\frac{g_{V}^{2}s^{4}}{48\pi^{2}F_{0}^{4}}\int_{4M_{\pi}^{2}}^{\infty} \frac{ds'}{s'(s'-s)}\sqrt{1-\frac{4M_{\pi}^{2}}{s'}}^{3} ,
\end{align}
where it is understood that real values of $s$ are approached from the upper complex plane for $s\in\lbrack 4M_{\pi}^{2},\infty\rbrack$. The expression in Eq.~(\ref{eq:Pirhopipi}) can be directly related to the 'bubble' type loop graph in Fig.~\ref{fig:loop}. The integral occuring here is given by
\begin{align}\label{eq:Jrhopipi}
\begin{split}
J_{\rho}^{\pi\pi}&=\int_{4M_{\pi}^{2}}^{\infty} \frac{ds'}{s'(s'-s)}\sqrt{1-\frac{4M_{\pi}^{2}}{s'}}^{3} \\
&= \frac{1}{s}\left(\frac{8}{3}-\frac{8M_{\pi}^{2}}{s}+2\sqrt{1-\frac{4M_{\pi}^{2}}{s}}^{3}\mathrm{Artanh}\left(-\frac{1}{\sqrt{1-\frac{4M_{\pi}^{2}}{s}}}\right)\right). 
\end{split}
\end{align}
For fixed nonzero $M_{\pi}$, the expansion of this integral in $s$ for $|s|<4M_{\pi}^{2}$ is given by
\begin{align}
J_{\rho}^{\pi\pi}=\frac{1}{10M_{\pi}^{2}}+\frac{s}{140M_{\pi}^{4}}+\ldots,
\end{align}
while the chiral expansion (for $4M_{\pi}^{2}<s$) is given by
\begin{align}\label{eq:Jrhopipiexp}
J_{\rho}^{\pi\pi}= \frac{1}{s}\left(\frac{8}{3}+\log\left(\frac{M_{\pi}^{2}}{s}\right)+i\pi-\frac{6M_{\pi}^{2}}{s}\left(1+\log\left(\frac{M_{\pi}^{2}}{s}\right)+i\pi\right)\right)+\mathcal{O}(M_{\pi}^{4}).
\end{align}
Matching the terms in Eq.~(\ref{eq:Pirhopipi}) to the expression for the dimensionally regularized loop graph of Fig.~\ref{fig:loop} and the first term in Eq.~(\ref{eq:PiRhoBbl}) ($d$ is the space-time dimension), 
\begin{align}
i\tilde{\Pi}_{\mu\nu}^{\pi\pi} &= i\left(g_{\mu\nu}-\frac{k_{\mu}k_{\nu}}{k^{2}}\right)\tilde{\Pi}_{\rho}^{\pi\pi},\\
\tilde{\Pi}_{\rho}^{\pi\pi} &= -\frac{4g_{V}^{2}}{F_{0}^{4}}s^{2}I_{A}^{\pi\pi}(s),\\
I_{A}^{\pi\pi}(s) &= \frac{1}{d-1}\left(\frac{1}{2}I_{\pi}-\frac{1}{4}(s-4M_{\pi}^{2})I_{\pi\pi}(s)\right),\label{eq:IApipi}
\end{align}
where the standard loop integrals $I_{\pi},I_{\pi\pi}$ can be found in app.~\ref{app:Loops}, we obtain simple expressions for the four subtraction constants:
\begin{align}
c_{0} &= c_{1} = 0,\\
c_{2} &= -\frac{g_{V}^{2}}{8\pi^{2}F_{0}^{4}}M_{\pi}^{2}\left(32\pi^{2}\bar{\lambda}+\log\left(\frac{M_{\pi}^{2}}{\mu^{2}}\right)\right),\\
c_{3} &= \frac{g_{V}^{2}}{48\pi^{2}F_{0}^{4}}\left(32\pi^{2}\bar{\lambda}+\log\left(\frac{M_{\pi}^{2}}{\mu^{2}}\right)+1\right).
\end{align}
The subtraction constant $c_{3}$ diverges logarithmically in the chiral limit, but this is counterbalanced by the infrared-divergent term in $J^{\pi\pi}_{\rho}$, see Eq.~(\ref{eq:Jrhopipiexp}).
Obviously, one needs counterterms of the form $s^{2}M_{\pi}^{2}$ and $s^{3}$, which are not yet present in the effective Lagrangian so far, to absorb the UV-divergent terms $\sim\bar{\lambda}$ in $c_{2,3}$. It is no problem in principle to construct counterterms with more derivatives acting on the vector fields, see e.g. \cite{Rosell:2004mn}. In effect, this would leave us with finite expressions for the subtraction constants,
\begin{align}
\bar{c}_{2} &= -\frac{g_{V}^{2}}{8\pi^{2}F_{0}^{4}}M_{\pi}^{2}\left(r_{2}(\mu)+\log\left(\frac{M_{\pi}^{2}}{\mu^{2}}\right)\right),\\
\bar{c}_{3} &= \frac{g_{V}^{2}}{48\pi^{2}F_{0}^{4}}\left(r_{3}(\mu)+\log\left(\frac{M_{\pi}^{2}}{\mu^{2}}\right)+1\right),
\end{align}
with unknown constants $r_{2,3}(\mu)$. In our counting, higher orders in $k^{2}=s$ are not suppressed, but, being interested in the resonance region, we can apply the reordering scheme of Eqs.~(\ref{eq:reorderingstep} - \ref{eq:newselfenergy}) to any polynomial in $s$, where $(s-M_{V}^{\star 2})$ is considered as a small quantity, in which one can expand the polynomial part of the self energies. UV-divergences of zeroth order in this quantity can then be absorbed in the parameters $M_{V}^{\star 2},b_{D}^{V}\ldots$, those of first order in the constants $f_{1}^{V\star},z_{D}^{V},z_{08}\ldots$ (see Eqs.~(\ref{eq:f1Rho} - \ref{eq:f108})), etc.
\section{Loop integrals}
\label{app:Loops}
Throughout this work we have used abbreviations for the appearing scalar integrals and in this section, we present the explicit expressions. The integrals containing only one propagator are given by 
\begin{align}
I_{M} &= \int\frac{d^{d}l}{(2\pi)^{d}}\frac{i}{l^{2}-M^{2}}=2M^{2}\bar{\lambda}+\frac{M^{2}}{16\pi^{2}}\log\left(\frac{M^{2}}{\mu^{2}}\right),\label{eq:tadpi}\\
I_{V} &= \int\frac{d^{d}l}{(2\pi)^{d}}\frac{i}{l^{2}-M_{V}^{2}}=2M_{V}^{2}\bar{\lambda}+\frac{M_{V}^{2}}{16\pi^{2}}\log\left(\frac{M_{V}^{2}}{\mu^{2}}\right).\label{eq:tadV}
\end{align}
The quantity $\bar{\lambda}$ contains the $1/\epsilon$-pole and some numerical constants. The subscript $M$ stands for the species of PGBs and the subscript $V$ for the vector particles. The quantity $M$ in the propagators represents any of the meson masses, i.e. $M_{\pi}$, $M_{K}$ and $M_{\eta}$. The scalar integral including two propagators can be split up into two parts:
\begin{align}\label{eq:IpiV}
I_{MV}(k^{2}\equiv s) = \int\frac{d^{d}l}{(2\pi)^{d}}\frac{i}{((k-l)^{2}-M_{V}^{2})(l^{2}-M^{2})} = I_{M V}(M_{V}^{2})-\frac{(s-M_{V}^{2})}{16\pi^{2}}J^{M V}(s),
\end{align}
where 
\begin{align}
\begin{split}
I_{M V}(M_{V}^{2}) &= 2\bar{\lambda}+\frac{1}{16\pi^{2}}\biggl(-1+\log\left(\frac{M_{V}^{2}}{\mu^{2}}\right) + \frac{M^{2}}{M_{V}^{2}}\log\left(\frac{M}{M_{V}}\right)\\ 
&\quad+ 2M\frac{\sqrt{4M_{V}^{2}-M^{2}}}{M_{V}^{2}}\mathrm{Arctan}\biggl(\frac{\sqrt{4M_{V}^{2}-M^{2}}}{2M_{V}+M}\biggr)\biggr)\\
&= 2\bar{\lambda}+\frac{1}{16\pi^{2}}\left(-1+\log\left(\frac{M_{V}^{2}}{\mu^{2}}\right)\right) + \frac{M}{16\pi M_{V}}\\ 
&\quad+ \frac{M^{2}}{16\pi^{2}M_{V}^{2}}\left(\log\left(\frac{M}{M_{V}}\right)-1\right)-\frac{M^{3}}{128\pi M_{V}^{3}}+\mathcal{O}(M^{4}),\label{eq:IpiVexp}
\end{split}
\end{align}
and $J^{MV}(s)$ is the finite function
\begin{align}
\begin{split}
J^{MV}(s) &= \int_{(M_{V}+M)^{2}}^{\infty} ds'\frac{\sqrt{(s'-(M_{V}+M)^{2})(s'-(M_{V}-M)^{2})}}{s'(s'-s)(s'-M_{V}^{2})}\\
&= \frac{M^{2}-M_{V}^{2}}{sM_{V}^{2}}\log\left(\frac{M}{M_{V}}\right) + \frac{4|\mathbf{q}|}{\sqrt{s}(s-M_{V}^{2})}\mathrm{Artanh}\biggl(\frac{2|\mathbf{q}|\sqrt{s}}{(M_{V}+M)^{2}-s}\biggr)\\ 
&\quad+ 2M\frac{\sqrt{4M_{V}^{2}-M^{2}}}{M_{V}^{2}(s-M_{V}^{2})}\mathrm{Arctan}\biggl(\frac{\sqrt{4M_{V}^{2}-M^{2}}}{2M_{V}+M}\biggr), 
\end{split}
\end{align}
where we have introduced the abbreviation 
\begin{align}\label{eq:qPiV}
|\mathbf{q}| = \frac{\sqrt{(s-(M_{V}+M)^{2})(s-(M_{V}-M)^{2})}}{2\sqrt{s}}.
\end{align}
We note that (for $M\not= 0$),
\begin{align}\label{eq:JpiVsfix}
\begin{split}
J^{MV}(M_{V}^{2}) &= -\frac{1}{M_{V}^{2}}\biggl(1+\log\left(\frac{M}{M_{V}}\right) - \frac{3\pi}{4}\frac{M}{M_{V}}+\frac{M^{2}}{M_{V}^{2}}\left(\frac{3}{4}-\log\left(\frac{M}{M_{V}}\right)\right)\\
&\quad+\mathcal{O}(M^{3})\biggr), 
\end{split}
\end{align}
while for $s\not= M_{V}^{2}$, the chiral expansion reads
\begin{align}
\begin{split}
J^{MV}(s) &= \frac{1}{s}\log\left(\frac{M_{V}^{2}}{M_{V}^{2}-s}\right) - \frac{\pi M}{M_{V}(M_{V}^{2}-s)} - \frac{M^{2}}{M_{V}^{2}}\frac{(3M_{V}^{2}-s)}{(M_{V}^{2}-s)^{2}}\log\left(\frac{M}{M_{V}}\right)\\ 
&+\frac{M^{2}}{M_{V}^{2}(M_{V}^{2}-s)^{2}}\left(2M_{V}^{2}-s-(s+M_{V}^{2})\frac{M_{V}^{2}}{s}\log\left(\frac{M_{V}^{2}}{M_{V}^{2}-s}\right)\right) \\
&\quad+ \frac{\pi}{8}\frac{M^{3}}{M_{V}^{3}(M_{V}^{2}-s)} + \mathcal{O}(M^{4}). 
\end{split}
\end{align}
The radius of convergence of the latter expansion vanishes as $s\rightarrow M_{V}^{2}$, in which case the expansion of Eq.~(\ref{eq:JpiVsfix}) must be used. The function $J^{MV}(s)$ diverges logarithmically when $s\rightarrow M_{V}^{2}$ and $M\rightarrow 0\,$.\\The basic integrals $I_{MM}$ ($I_{\pi\pi}$, $I_{\bar{K}\pi}\ldots$) can be found directly from Eq.~(\ref{eq:IpiV}) by replacing $M_{V}$ by the corresponding mass of $\pi,K$ or $\eta$ (of course, the chiral expansion is completely different in that case).\\Furthermore, in the main text we have used the integral $I_A^{MV}$, which is given by the following expression:
\begin{align}
\begin{split}\label{eq:IMVA}
I_A^{MV}&=\frac{1}{4s(d-1)}\biggl((4sM^2-(s+M^2-M_V^2)^2)I_{MV}+(s+M^2-M_V^2)I_{M}\\
&\quad+(s-M^2+M_V^2)I_{V}\biggr) 
\end{split}
\end{align}
The integral $I_A^{MM}$ can be obtained by replacing $M_V$ with $M$ in Eq. \eqref{eq:IMVA} (see e.g. Eq.~(\ref{eq:IApipi})). The integrals $I_{A}^{MS}$ are of course of the same form as $I_{A}^{MV}$, with $M_{V}$ replaced by the mass of the singlet vector meson, $M_{S}$. In the chiral limit, the integrals $I_{A}^{MM}$, $I_{A}^{MV}$ are given by
\begin{align}
I_{A}^{MM}(s) &\rightarrow \frac{s}{192\pi^{2}}\left(\frac{5}{3} + \log\left(-\frac{\mu^{2}}{s}\right) -32\pi^{2}\bar{\lambda}\right) \label{eq:IApipichlim}\,\\
\begin{split}
I_{A}^{MV}(s) &\rightarrow \frac{1}{192\pi^{2}s}\biggl((3M_{V}^{2}s-s^{2})\left(32\pi^{2}\bar{\lambda} + \log\left(\frac{M_{V}^{2}}{\mu^{2}}\right)\right)\\  
&\quad+ \frac{(s-M_{V}^{2})^{3}}{s}\log\left(\frac{M_{V}^{2}}{M_{V}^{2}-s}\right)+ \frac{1}{3}\left(5s^{2}-12M_{V}^{2}s+3M_{V}^{4}\right)\biggr) .\label{eq:IApiVchlim} 
\end{split}
\end{align}
The chiral expansion of $I_{A}^{MV}(s)$ is found to read, for $s\not= M_{V}^{2}$,
\begin{align}
\begin{split}
I_{A}^{MV}(s) &= \frac{1}{192\pi^{2}s}\biggl((3M_{V}^{2}s-s^{2})\left(32\pi^{2}\bar{\lambda} + \log\left(\frac{M_{V}^{2}}{\mu^{2}}\right)\right)\\
&\quad + \frac{(s-M_{V}^{2})^{3}}{s}\log\left(\frac{M_{V}^{2}}{M_{V}^{2}-s}\right)+ \frac{1}{3}\left(5s^{2}-12M_{V}^{2}s+3M_{V}^{4}\right)\\ 
&+ \frac{3M^2}{s}\left(s^{2}\left(32\pi^{2}\bar{\lambda} + \log\left(\frac{M_{V}^{2}}{\mu^{2}}\right)\right) + (M_{V}^{4}-s^{2})\log\left(\frac{M_{V}^{2}}{M_{V}^{2}-s}\right) \right.\\
&\quad\left.- s(s+M_{V}^{2})\right)+ \mathcal{O}(M^{4}) \biggr) .\label{eq:IApiVchexp} 
\end{split}
\end{align}
\acknowledgments{This work was supported by the Deutsche Forschungsgemeinschaft SFB/Transregio 55.}
\bibliographystyle{apsrev}

\end{document}